\documentclass[pdflatex,sn-mathphys-num]{sn-jnl}

\usepackage{graphicx}%
\usepackage{multirow}%
\usepackage{amsmath,amssymb,amsfonts}%
\usepackage{amsthm}%
\usepackage{mathrsfs}%
\usepackage[title]{appendix}%
\usepackage{xcolor}%
\usepackage{textcomp}%
\usepackage{manyfoot}%
\usepackage{booktabs}%
\usepackage{algorithm}%
\usepackage{algorithmicx}%
\usepackage{algpseudocode}%
\usepackage{listings}%
\usepackage{comment}
\usepackage{todonotes}
\usepackage{lineno} 
\usepackage[none]{hyphenat}
\nolinenumbers
\newcommand{\cotwo}{{CO${}_2$}}

\usepackage{pifont}
\theoremstyle{thmstyleone}%
\theoremstyle{thmstyletwo}%

\theoremstyle{thmstylethree}%

\begin{document}

\title[Article Title]{Examining Fast Radiatively Driven Responses Using Machine-Learning Weather Emulators}



\author*[1,2]{\fnm{Ankur} \sur{Mahesh}}\email{amahesh@lbl.gov}
\equalcont{These authors contributed equally to this work.}

\author*[1,2]{\fnm{William D.}\sur{Collins}}\email{wdcollins@lbl.gov}
\equalcont{These authors contributed equally to this work.}

\author[3,2]{\fnm{Travis A.} \sur{O'Brien}}
\author[3]{\fnm{Paul B.} \sur{Goddard}}
\author[3]{\fnm{Sinclaire} \sur{Zebaze}}
\author[4]{\fnm{Shashank} \sur{Subramanian}}
\author[5]{\fnm{James P.C.} \sur{Duncan}} 
\author[5]{\fnm{Oliver} \sur{Watt-Meyer}}
\author[6]{\fnm{Boris} \sur{Bonev}}
\author[6]{\fnm{Thorsten} \sur{Kurth}}
\author[6]{\fnm{Karthik} \sur{Kashinath}}
\author[6]{\fnm{Michael S.} \sur{Pritchard}}
\author[7]{\fnm{Da}\sur{Yang}}

\affil[1]{\orgdiv{Earth and Planetary Science}, \orgname{University of California, Berkeley}, \orgaddress{
\city{Berkeley}, 
\state{CA}, \country{USA}}}

\affil[2]{\orgdiv{Earth and Environmental Sciences}, \orgname{Lawrence Berkeley National Laboratory}, \orgaddress{
\city{Berkeley}, 
\state{CA}, \country{USA}}}

\affil[3]{\orgdiv{Earth and Atmospheric Sciences}, \orgname{Indiana University}, \orgaddress{
\city{Bloomington}, 
\state{IN}, \country{USA}}}

\affil[4]{\orgdiv{Computational Sciences}, \orgname{Lawrence Berkeley National Laboratory}, \orgaddress{
\city{Berkeley}, 
\state{CA}, \country{USA}}}

\affil[5]{\orgname{Allen Institute for Artificial Intelligence (Ai2)}, \orgaddress{
\city{Seattle}, 
\state{WA}, \country{USA}}}

\affil[6]{\orgname{NVIDIA Corporation}, \orgaddress{
\city{Santa Clara}, 
\state{CA}, \country{USA}}}

\affil[7]{\orgdiv{Department of Geophysics}, \orgname{Stanford University}, \orgaddress{
\city{Stanford},
\state{CA}, \country{USA}}}

\abstract{The response of the climate system to increased greenhouse gases and other radiative perturbations is governed by a combination of fast and slow responses.  Slow responses are typically activated in response to changes in ocean temperatures on decadal timescales and manifest as changes in climatic state with no recent historical analogue.  However, fast responses are activated in response to rapid atmospheric physical processes on weekly timescales, and they are already operative in the present-day climate.  This distinction implies that the physics of fast radiatively driven responses is present in the historical meteorological reanalyses used to train many recent successful machine-learning-based (ML) emulators of weather and climate.  In addition, these responses are functional under the historical boundary conditions pertaining to the top-of-atmosphere radiative balance and sea-surface temperatures.  Together, these factors imply that we can use historically trained ML weather emulators to study the response of radiative-convective equilibrium (RCE), and hence the global hydrological cycle, to perturbations in \cotwo\  and other well-mixed greenhouse gases.  Without retraining on prospective Earth system conditions, we use ML weather emulators to quantify the fast precipitation response to reduced and elevated \cotwo\ concentrations with no recent historical precedent. We show that the responses from historically trained emulators agree with those produced by full-physics Earth System Models (ESMs). In conclusion, we discuss the prospects for and advantages from using ESMs and ML emulators to study fast processes in global climate.}

\keywords{atmospheric science, machine learning}



\maketitle

\section{Introduction}\label{sec1}
In climate science, a central challenge is to predict the future evolution of the climate system given only observations available up to the present. The response to increased greenhouse gases (GHGs) is governed by a combination of fast and slow responses. Slow responses are activated in response to decadal changes in ocean temperatures and introduce changes in climatic state with no recent historical analogue. In contrast, fast responses emerge from atmospheric radiative and convective processes operating on weekly timescales, and these are already fully operative in the present-day climate. 

Radiative–convective equilibrium (RCE) is a key process in Earth’s troposphere that governs these fast responses.  RCE brings the troposphere into a global, time-mean thermal equilibrium by mostly balancing atmospheric radiative cooling with latent heating released when water vapor is lifted by convection and condenses to form clouds \citep{Ramanathan1989, Jeevanjee2018}.  Based on the ratio of the global atmospheric water content to the global mean rainfall rate, RCE adjusts on a characteristic timescale of roughly one week, which is orders of magnitude faster than the thermal response of the upper ocean. Radiation itself responds almost instantly to radiatively active agents in the Earth's atmosphere such as GHGs.  Hence, the processes governing the response of RCE to an instantaneous pulse of GHGs are very fast compared to the slow thermally driven feedbacks in the climate system. These fast adjustments, first identified in simulations of instantaneous \cotwo\  forcing, have since been confirmed across multi-model ensembles and play a central role in shaping the hydrological response to climate forcing \citep{Stjern2023, Samset2016, Andrews2008,Gregory2008,Andrews2009,Andrews2010, Kamae2015, Tian2017, Douville2020, Richardson2016, Dong2009}.

\begin{figure}
    \centering
    \includegraphics[width=\linewidth]{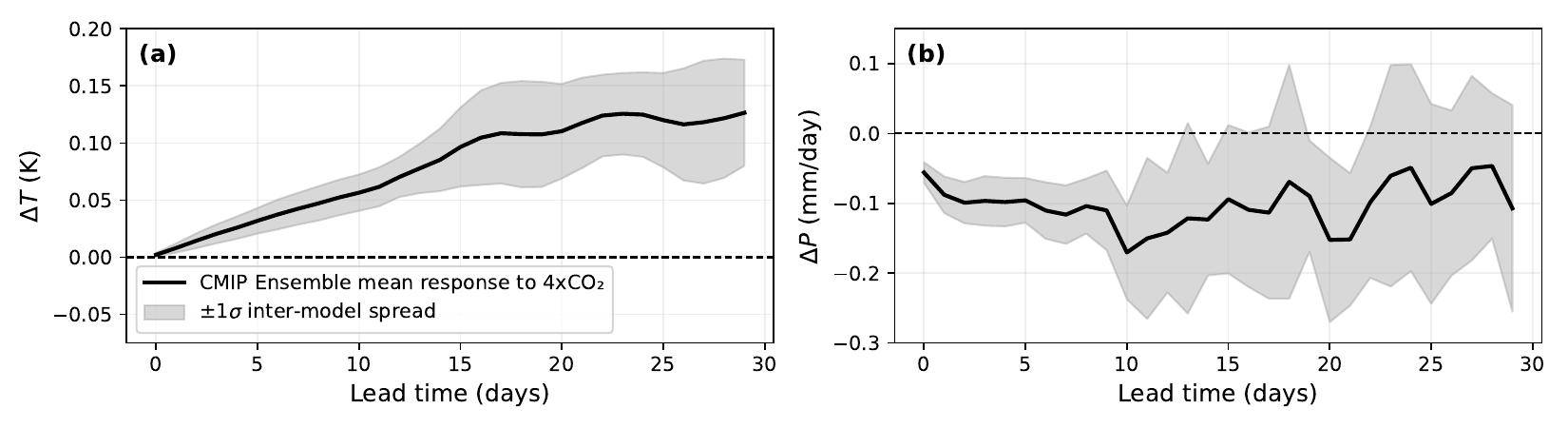}
    \caption{\textbf{Response of CMIP Models to Abrupt \cotwo\  Quadrupling.} The response of 8 ESMs to the abrupt4xCO2  experiment from the Coupled Model Intercomparison Project (CMIP) v6 \citep{Eyring2016}.  Lead time is shown as days since \cotwo\  was quadrupled. Listed in Table~\ref{table:cmip_table}, models were selected based on availability of data at daily resolution and sufficient metadata to determine the branch points of the instantaneous quadrupling simulations from the corresponding pre-industrial control simulations.  Panel~(a) shows the response of ocean surface temperature, and panel~(b) shows the response of precipitation. For precipitation and temperature, the $\Delta$ refers to the global-mean differences between the abrupt4xCO2 simulations and the pre-industrial control simulations from which these were branched.  Shading denotes $\pm1$ standard deviation across the multi-model ensemble.}
    \label{fig:cmip_ensemble}
\end{figure}

These processes can be illustrated using a representative subset of the physics-based Earth System Models (ESMs) submitted to the Coupled Model Intercomparison Project Phase 6 (CMIP6) \citep{Eyring2016}. In the abrupt4xCO2  CMIP6 experiment, an instantaneous quadrupling of atmospheric \cotwo\  from pre-industrial concentrations reduces atmospheric radiative cooling \citep{Jeevanjee2018}.    Figure~\ref{fig:cmip_ensemble}a shows that global-mean sea surface temperatures (SSTs) in these ESMs change relative to the pre-industrial control simulations over the course of one month by only approximately 0.1 $^{\circ}$C, a small fraction of the equilibrated response in SST.  Therefore, during the first month,  the upper and lower boundary conditions on the Earth's atmosphere, i.e., the solar insolation and sea-surface temperatures, remain approximately the same as the boundary conditions in the preindustrial simulations. On $\sim$10-day timescales, the hydrological cycle responds to the reduced radiative cooling with a corresponding reduction in latent heating from condensation (and hence precipitation) \cite{Stjern2023}.  Figure~\ref{fig:cmip_ensemble}b shows this fast response in the CMIP6 multi-model ensemble. 

Here, we test whether machine learning (ML) emulators of weather and climate can reproduce these fast adjustments of RCE without being trained on simulations with elevated \cotwo. Physics-based ESMs approach climate prediction by encoding fundamental laws of physics, chemistry, and biogeochemistry, approximating subgrid-scale physical processes, and extrapolating into future climate states. On the other hand, ML emulators are trained to emulate the time evolution of the atmosphere directly from historical reanalyses \citep{paper_173, Bi2023, Lam2023, paper_171, Bonev2023}.  While they can be expensive to train, ML emulators are significantly faster and more energy-efficient during inference than their physics-based counterparts \citep{Pathak2022}.  ML emulators also maintain similar accuracy on weather and climate timescales \citep{Bonev2025, Watt-meyer2023, Bi2023, Lam2023, CresswellClay2025, Chapman2025}.  Here we use the Allen Institute for Artificial Intelligence (Ai2) Climate Emulator (ACE) \cite{Duncan2024}, an ML emulator that autoregressively predicts the three-dimensional atmospheric state with a 6-hour timestep and 1-degree horizontal resolution (Section~\ref{ssec:ACE}).  We use the version of ACE trained on 40 years of simulations from a physics-based atmosphere model called the Energy Exascale Earth System Model Atmosphere Model (EAMv2) \cite{Duncan2024, Golaz2022}. In these simulations, EAMv2 uses an annually repeating cycle of climatological SSTs (2005–2014 average) as its boundary condition, and it has fixed greenhouse gas emissions from the year 2010 (the ``perpetual 2010,'' or ``F2010,'' scenario).


Using ML emulators trained on this historical climate, we compute the fast precipitation responses to large variations in \cotwo\ unprecedented in recent geological history. We test this capability using a full radiative transfer model and with an idealized experiment designed to rigorously probe the models’ learned physics.  We present architectural modifications that are crucial to enable ML emulators to represent the column physics of RCE in the training dataset.  With these changes, ML emulators are in strong agreement with full-physics ESMs. These results show that fast radiative–convective adjustments (one of the most important components of the global hydrological response \citep{Andrews2010, Samset2016}) can be accurately captured by data-driven emulators trained solely on the historical climate. 


\section{Fast Radiatively Driven Responses to \cotwo\  in ML Emulators}

\subsection{Changing \cotwo\ in the ML Emulator with Physics-based Radiative Transfer}
\label{ssec:ChangingCO2}
\begin{figure}
    \centering
    \includegraphics[width=\linewidth]{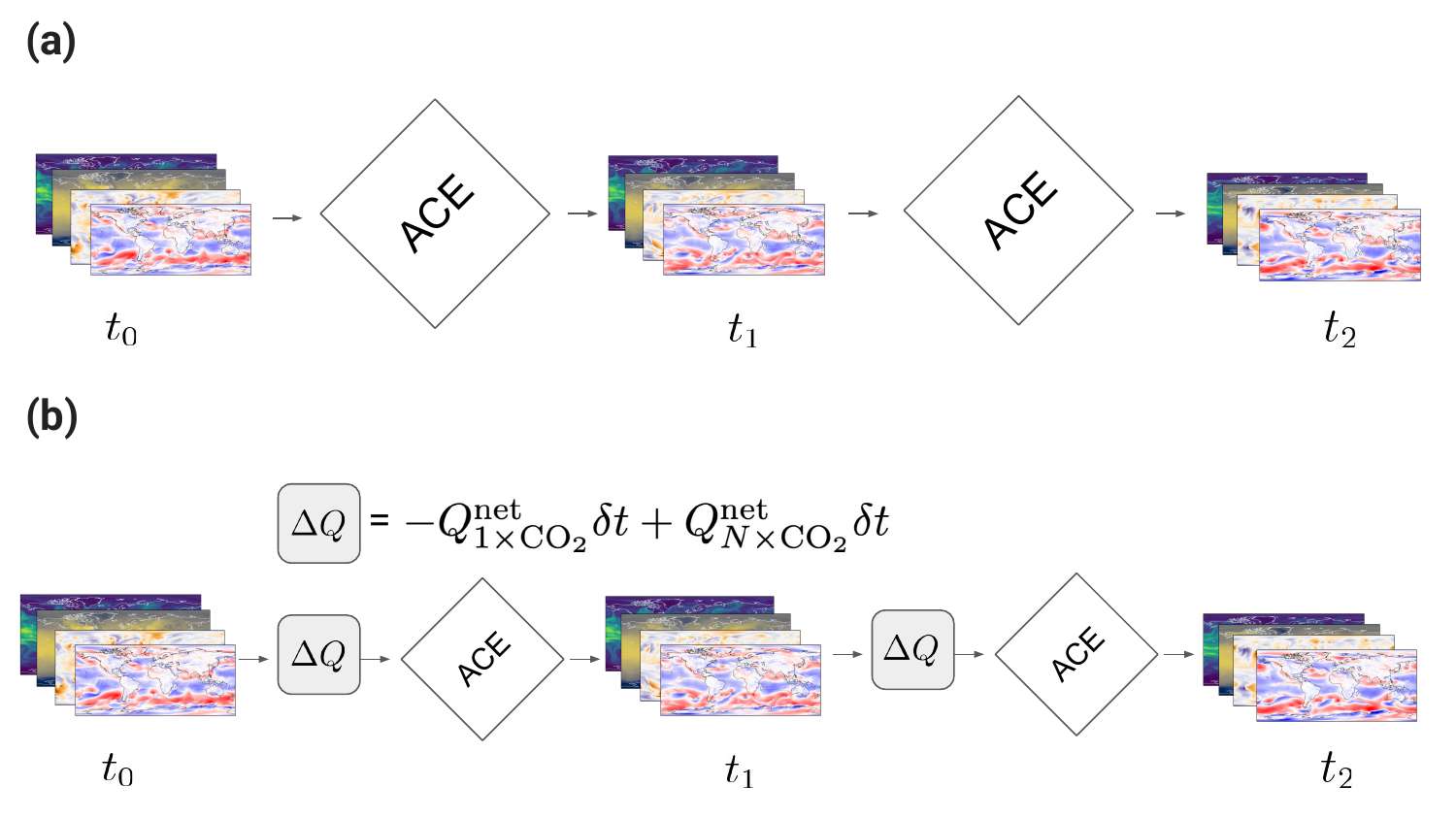}
    \caption{\textbf{Diagram of ACE-RRTMG Rollout}. Panel~(a) shows the standard rollout of ACE with no perturbations for \cotwo\ applied. The instantaneous model states are depicted at times $t_0 , t_1, \text{and}\  t_2$ with stylized maps depicting various prognosed variables. Panel~(b) shows ACE-RRTMG with modifications to simulate the effects of different levels of \cotwo.  $Q$ is the net clear-sky radiative heating [${}^\circ$K], and $Q^{\text{net}}$ is the sum of the clear-sky longwave and shortwave heating rates [${}^\circ$K/s]. At each timestep, RRTMG is used to calculate $\Delta Q$, the heating  perturbation due to an instantaneous change in \cotwo\  from $1\times$ to $N\times$ its concentration at 2010 CE.  The heating rate perturbations are applied to the temperature fields at each pressure level in the ACE rollout.}
    \label{fig:rollout_modification}
\end{figure}

We conduct 1-month forecasts with ACE assuming different levels of \cotwo. In 2010 CE (Common Era), the CO$_2$ concentration was 388 ppm, and we refer to this level as ``1 $\times$ CO$_2$.'' To change the \cotwo\  level, we couple ACE to the Rapid Radiative Transfer Model for GCMs (RRTMG) (Section~\ref{sec:methods_modified_ace_rollout}), a physics-based radiative transfer package \citep{paper_057}.  For a given atmospheric state, RRTMG calculates radiative heating rates based on a specified \cotwo\  concentration. At each forecast step, we subtract the clear-sky 1$\times$\cotwo\  radiative heating rate and add the clear-sky heating rate assuming N$\times$\cotwo\  (Figure~\ref{fig:rollout_modification}). This yields the temperature perturbations to simulate the rollout in an N$\times$\cotwo\ world instead of a 1$\times$\cotwo\ world. We refer to this rollout as ``ACE-RRTMG.'' We perform the analogous experiment using EAMv2 (Section~\ref{ssec:EAMv2}).  EAMv2 also uses RRTMG with the same shortwave and longwave correlated-k bands and the same versions of the optical parameters for water vapor and greenhouse gases employed in ACE.  We utilize perturbations for clear-sky rather than all-sky conditions because ACE does not emulate cloud fields directly and instead uses specific total water (the sum of water vapor and condensates) as its moisture variable.

ACE-RRTMG is a hybrid modeling approach in which ACE predicts the atmospheric time evolution with ML while retaining a physics-based radiative transfer scheme through RRTMG. The physics-based radiative transfer code enables reliably emulating the effects of changing \cotwo\  during rollout even though the version of ACE we employ does not explicitly or implicitly input \cotwo.  This hybrid approach retains physics-based control over the external forcing while leveraging ML to simulate the atmospheric time evolution, and it permits interactive modulation of the clear-sky radiative feedback to elevated greenhouse gas while maintaining consistency with the locally evolving state vector. By subjecting data-driven emulators to standard physical radiative perturbation tests such as fixed-SST \cotwo\ experiments, we directly assess whether they reproduce established physical responses.

\begin{figure}
    \centering
    \includegraphics[width=\linewidth]{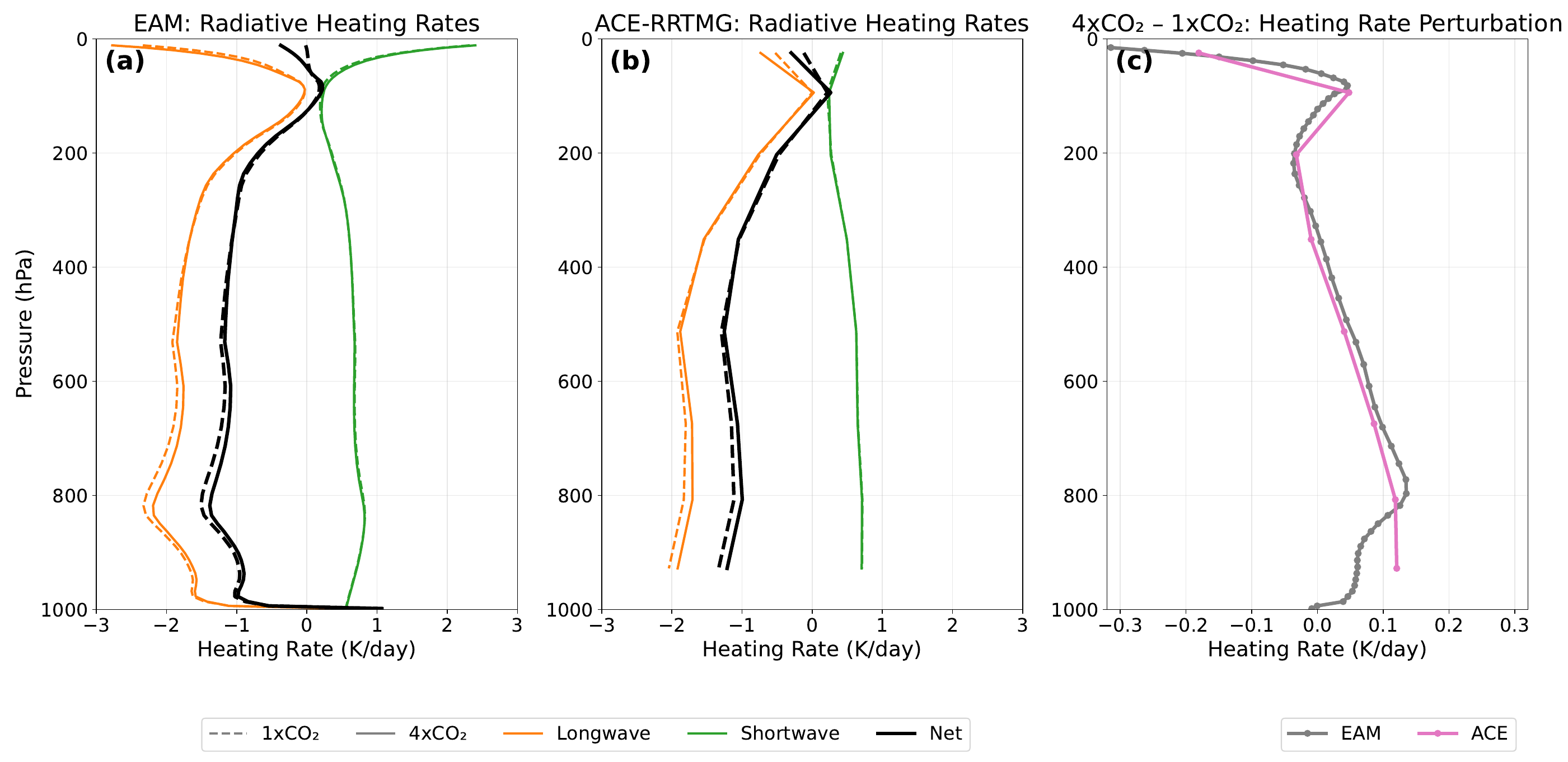}
    \caption{\textbf{Heating Rate Perturbations from 1$\times$\cotwo\  \& 4$\times$\cotwo}. The longwave, shortwave, and net heating rate perturbations are shown for the E3SM Atmosphere Model EAMv2 in panel~(a) and ACE coupled to RRTMG in panel~(b).  The net heating rate perturbation (4$\times$\cotwo\  - 1$\times$\cotwo) is shown for EAMv2 and ACE in~(c).  Note the different scales in the x axis in (c) compared to (a) and (b).}
    \label{fig:heating_rate_perturbations}
\end{figure}

ACE-RRTMG successfully replicates the vertical profile of radiative heating from the EAMv2 simulation. Figures~\ref{fig:heating_rate_perturbations}a and~\ref{fig:heating_rate_perturbations}b show the longwave, shortwave, and net heating rates from EAMv2 and ACE-RRTMG.  The net heating rates applied to ACE-RRTMG (Figure~\ref{fig:rollout_modification}) are the sums of the longwave and shortwave components.  ACE utilizes eight vertical model levels and hence has a coarser vertical resolution than EAMv2 discretized onto 72 model levels (Table~\ref{table:hybrid_levels}).  After accounting for this difference, the heating rate perturbations assuming elevated \cotwo\  are similar in the two models (Figure~\ref{fig:heating_rate_perturbations}c).  In the troposphere, the heating rate perturbation from increased \cotwo\  is approximately 0.1 K/day.  This perturbation is small compared to the total radiative tendencies themselves (Figures~\ref{fig:heating_rate_perturbations}a and \ref{fig:heating_rate_perturbations}b).  It is also significantly smaller than day-to-day variations in weather, e.g. from synoptic-scale storms.  Therefore, we hypothesize that ACE can successfully simulate the response of the hydrological cycle to these perturbations.

\subsection{Global Mean Fast Responses}

\begin{figure}
    \centering
    \includegraphics[width=\linewidth]{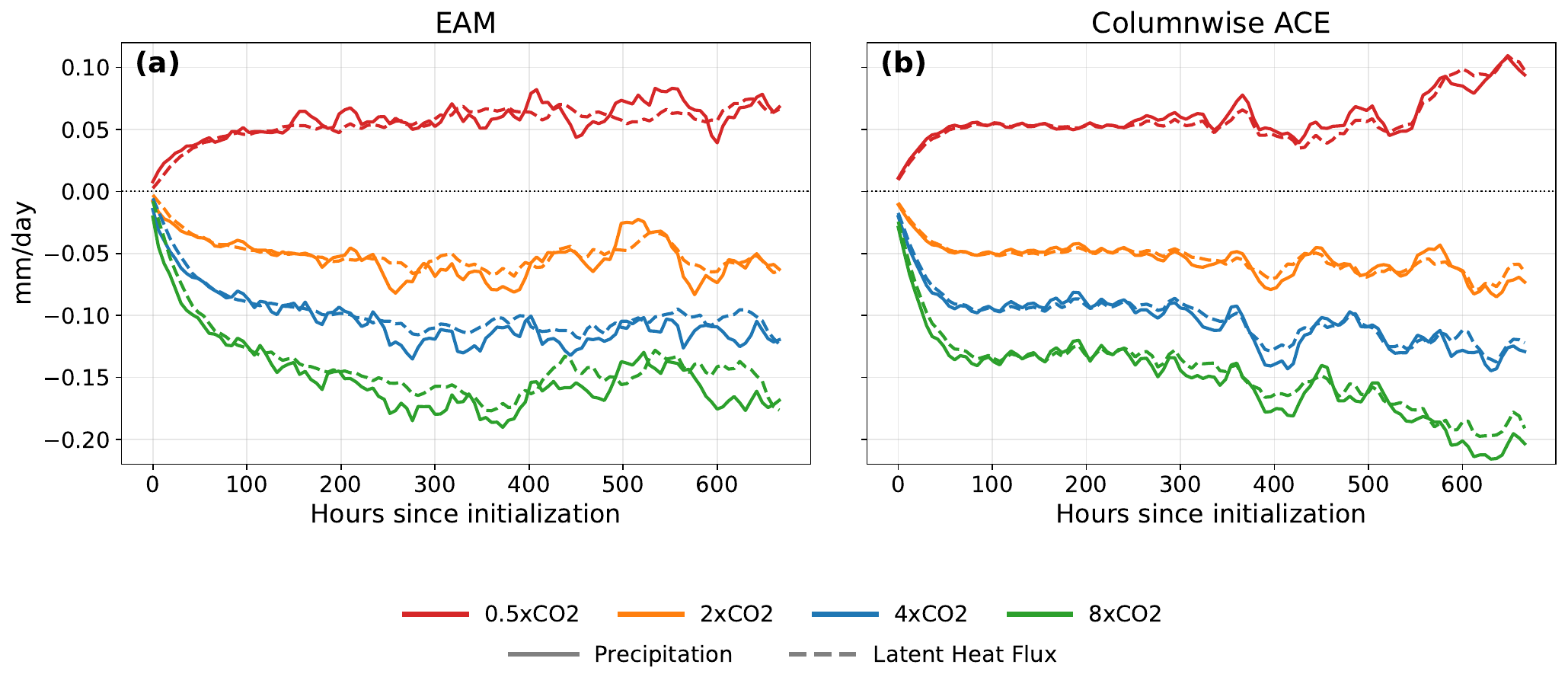}
    \caption{\textbf{Global Mean Fast Response to Instantaneous Changes in \cotwo\ Concentrations}. Panel~(a): the fast response to increased \cotwo\  from the E3SM Atmosphere Model; panel~(b): the corresponding response from our modified ACE architecture including column diagnosis and prognosis.  See Section~\ref{ssec:ace_modification} and Figure~\ref{fig:architecture_diagram} for a description of these modifications.  Colors denote the response to instantaneous multiplicative changes in \cotwo\  concentration relative to a $1\times$ baseline. Dashed lines denote latent heat flux, and solid lines denote precipitation.  Responses are calculated using  96-member ensembles of 1-month-long simulations from EAMv2 and columnwise ACE.}
    \label{fig:global_mean_response}
\end{figure}

In response to increased \cotwo, global mean precipitation and latent heat flux decrease to maintain RCE in EAMv2 (Figure~\ref{fig:global_mean_response}a), consistent with other CMIP6 models (Figure~\ref{fig:cmip_ensemble}).  Conversely, if \cotwo\  is decreased, then precipitation and latent heat flux increase.  Larger \cotwo\ changes imply larger precipitation and evaporation changes, consistent with the logarithmic radiative forcing of \cotwo\  \citep{Romps2022}.  These responses are calculated from averages of 96-member ensembles, with each member initialized at the beginning of each month over a span of 8 years (see Sections~\ref{ssec:EAMv2} and ~\ref{sec:hyperparameter_tuning}).

We introduce a modified architecture of ACE that diagnoses precipitation and latent heat flux solely based on the local vertical profiles of the atmospheric state.  See Figure~\ref{fig:architecture_diagram} and Section~\ref{sec:ace_modifications} for a description of all changes made to the ACE version presented in \cite{Duncan2024}. This choice is motivated by the column-local abstraction in physics-based models \cite{Balaji2022}, which also diagnose precipitation using the vertical profile of the atmospheric state. By default, the Spherical Fourier Neural Operators in ACE mix all global information to predict its diagnostic outputs including precipitation (Table~\ref{table:variable_list}). Empirical tests led us to speculate that this mixing can lead to degraded performance when the atmosphere is perturbed with a globally uniform field  (Section~\ref{ssec:criteria_for_success} and Figures~\ref{fig:global_mean_response_defaultace} and~\ref{fig:spatial_patterns_defaultace}). Figure~\ref{fig:heating_rate_perturbation_vis} shows that the \cotwo\ heating rate perturbations of $\mathcal{O}(0.1 K/ \text{day})$ are globally uniform but are also small compared to climatological radiative cooling rates. Although the training dataset contains variability larger than $\mathcal{O}(0.1 K/ \text{day})$ due to synoptic-scale weather patterns, these variations do not occur uniformly across the globe. The default ACE architecture mixes global information to diagnose precipitation, so it does not respond correctly to a global perturbation. However, with a column-local diagnosis approach, ACE can correctly respond to small global perturbations, such as those from \cotwo. 

Using the exact same training data as the default ACE in \cite{Duncan2024}, the columnwise ACE correctly simulates the response of the hydrological cycle to increased \cotwo. As illustrated in Figure~\ref{fig:global_mean_response}b, the modified ACE accurately captures both the signs and magnitudes of the changes in precipitation and latent heat flux subject to \cotwo\  forcing.  Due to chaotic weather variability, the ensemble spread in the fast \cotwo\ response increases with lead time. Shown in Figure~\ref{fig:ensemble_spread_demo}, the spread is substantial by the end of the simulation, exceeding any apparent, likely spurious, late-time trend in the ensemble mean. To verify that ACE has learned the correct response for the correct reasons, we evaluate the performance of ACE to an idealized perturbation for which the precipitation response can be estimated directly from first principles without resorting to a model (see Section~\ref{ssec:idealized_test}). Figure~\ref{fig:idealized_test_ace} shows that the column-local ACE passes this idealized test.

\subsection{Spatial Patterns of the Hydrological Cycle Response}

\begin{figure}
    \centering
    \includegraphics[width=\linewidth]{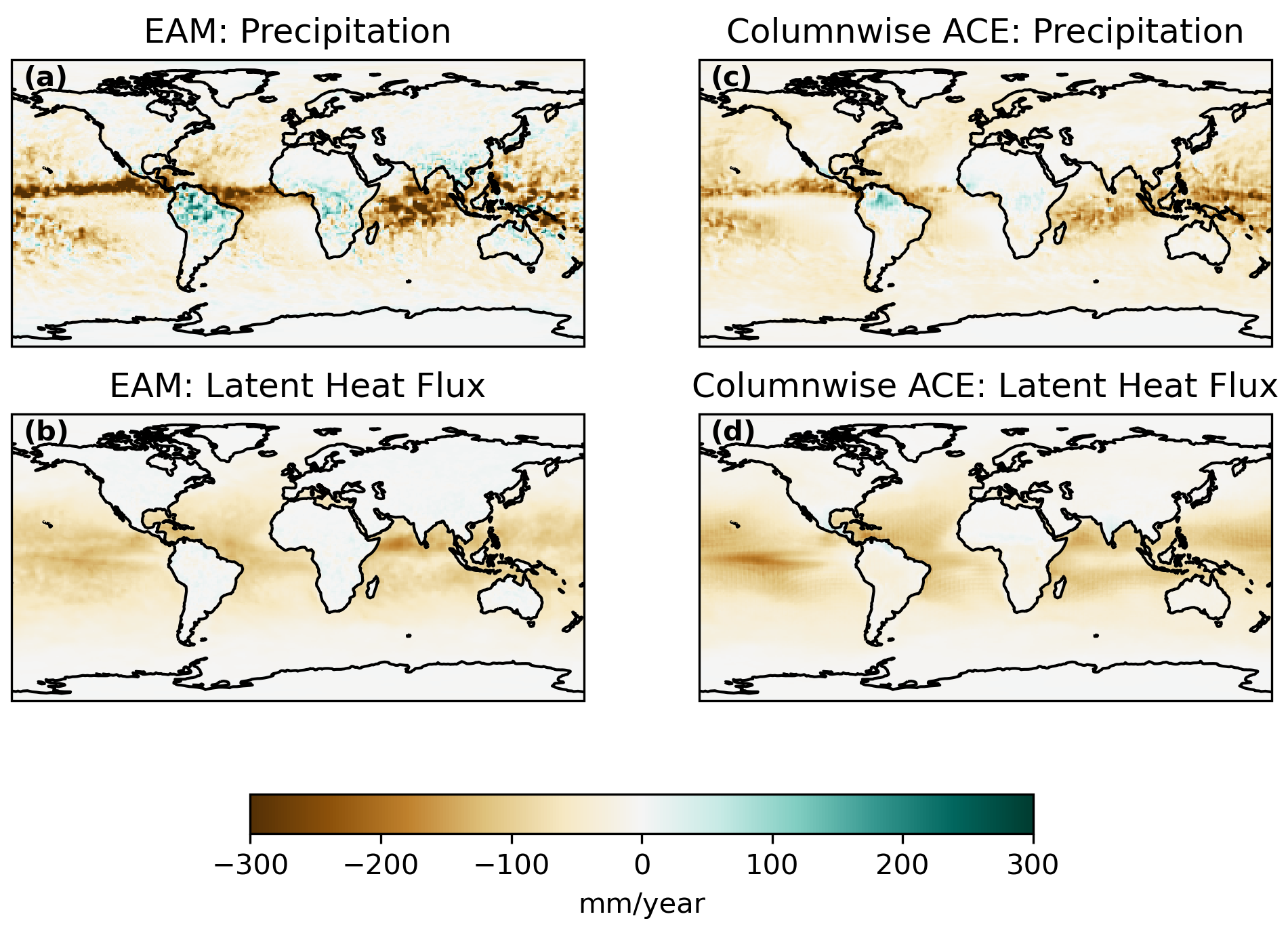}
    \caption{\textbf{Spatial Pattern of Precipitation and Latent Heat Flux Fast Response to 8$\times$\cotwo}. Spatial patterns are calculated from the first seven days after \cotwo\  is instantaneously octupled. Panels~(a) and~(b): response from the physics-based E3SM Atmosphere Model; panels~(c) and~(d): response from our modified ACE architecture  with column diagnostics and prognostics.  Panels~(a) and~(c): spatial pattern of the response of precipitation; panels~(b) and~(d): corresponding response of latent heat flux converted to equivalent precipitation units.}
    \label{fig:spatial_patterns}
\end{figure}

Figure~\ref{fig:spatial_patterns} shows the spatial patterns of ACE's response during the first 7 days of the rollout. We focus on this timescale because it minimizes the impact of chaotic weather noise, which tends to dominate after 14 days due to the limit of predictability \cite{LORENZ1969, charney1966feasibility}. Additionally, a 7-day timeframe is comparable to the turnover time of water vapor in the atmosphere. (The patterns for all 28 days of the rollout are shown in Figure~\ref{fig:spatial_patterns_alldays}.) The columnwise ACE produces spatial patterns that are in broad agreement with EAMv2.  Even though the temperature perturbations are largely uniform in the tropics and midlatitudes (Figure~\ref{fig:heating_rate_perturbation_vis}), the columnwise ACE correctly simulates that the response of precipitation and latent heat flux has its own spatial structure, unique to the climatological characteristics of those variables.

However, compared to ACE, the EAM response is muted at individual grid cells.  One reason for this muting is that many deterministic ML emulators display blurred predictions compared to their training dataset due to the use of a large timestep and a mean-squared error loss function \citep{Kochkov2024, Keisler2022, Brenowitz2025, Lang2024}.  Both the columnwise and default ACE versions display a slightly degraded precipitation spectrum compared to EAMv2 (Figure~\ref{fig:precip_hist_spectra}). A second reason for the muting is the different land-sea contrasts in the two models.  EAMv2 exhibits a larger increase in terrestrial precipitation than ACE in response to elevated \cotwo, particularly over South America, Africa, India, and China (Figures~\ref{fig:spatial_patterns}a and c).  Since the columnwise ACE and EAMv2 have similar global mean responses, the land-sea differences between these models imply that the columnwise ACE must exhibit a muted oceanic precipitation response.  Improving the spatial patterns of the EAMv2 response (e.g., with a land model integrated or a probabilistic loss function \cite{Perkins2025}) is an area of future model development.

\section{Discussion}
\label{sec:Conclusions}

We show that data-driven ML emulators of the climate system can predict the fast responses of the global hydrological cycle to \cotwo.  We also introduce an idealized perturbation test, in which the precipitation response can be calculated directly (Figure~\ref{fig:idealized_test_ace} and Section~\ref{ssec:idealized_test}). The \cotwo\ experiment requires additional programmer effort and compute time to integrate a full radiative transfer code (RRTMG) into the ML emulator rollout. In contrast, the idealized test provides a computationally inexpensive benchmark to assess the representation of RCE and fast responses; it can be readily applied to other ML climate emulators to improve model development.  Together, these tests show that it is possible, at least in principle, to substitute ML emulators in place of traditional ESMs in numerical experiments with fixed sea-surface temperatures.  Such experiments are a key component of international efforts such as the Coupled Model Intercomparison Project Phase 7 (CMIP7) \cite{Dunne2024}.  This work paves the way to extend these tests to other radiative forcing agents, such as other GHGs and short-lived radiative agents like black carbon \cite{Andrews2010}. 

In full-physics general circulation models, subgrid-scale phenomena and diagnostic variables (including precipitation) are parameterized using a column abstraction. These processes are calculated only based on the state variables of a single vertical atmospheric column. This abstraction is thought to be a shortcoming because some phenomena like tropical convective organization occur on scales that are larger than a single column \cite{Balaji2022}. While the default ACE is fully non-local and unconstrained by this abstraction, we find that reintroducing the column abstraction is a way to capture the correct physical response in ACE.  Like most ML emulators, ACE uses a large timestep (6 hours), which is longer than physics-based models of comparable resolution.  Given the large timestep, we predict precipitation using a 3 $\times$ 3 grid-cell patch to account for some interaction with neighboring columns at 6-hour timescales.  With an approximately 100 km horizontal grid resolution and a 6-hour timestep, the near-column diagnostic model can account for phenomena propagating at speeds up to 83 m/s in its precipitation diagnosis \cite{Karlbauer2024}. We find that the 3 $\times$ 3 receptive field is sufficient to predict the correct precipitation distribution (Figure~\ref{fig:precip_hist_spectra}), a realistic time-longitude precipitation evolution (Figure~\ref{fig:precip_hovmoller}), and low bias in 10-year rollouts (Figure~\ref{fig:spatial_bias_maps}).  As the field of ML-based climate emulation develops, an open question concerns what architectural inductive biases, like the column abstraction used here, are necessary.  This will inform the degree to which ML model architectures contain local and non-local components, which our findings here emphasize is a very important design choice \cite{Lam2023, Pereira2026}.



ML emulators trained on reanalysis datasets have significantly less biased simulations of the current climate compared to physics-based models \citep{Watt-meyer2025, Kochkov2024, Yuval2026, CresswellClay2025}.  On the other hand, ML emulators trained on physics-based models inherit their biases.  While this is considered a disadvantage, the experiment here shows the value of this emulation.  For climate scenarios that have no observations, ML emulators can be tested against a physics-based model, and the correct response can be verified \cite{Clark2025}. As we find with the need for column-local modifications, the correct response does not necessarily appear in all ML architectures, so this is a critical workflow step for ML validation.  We propose that such perfect model tests should be a prerequisite before using reanalysis-trained ML emulators to simulate climate change.  



ACE-RRTMG is a new hybrid modeling framework. Existing hybrid ML emulators combine a physics-based dynamical core with machine-learned parameterizations \cite{Kochkov2024}. ACE-RRTMG instead adopts the complementary strategy: it retains a physics-based radiation scheme while using ML for all other model components. This design enables physics-based experiments and sensitivity tests, which cannot be performed with purely ML-based frameworks. For example, varying atmospheric implied \cotwo\ concentrations would not be possible in ACE alone, but it becomes feasible through the inclusion of RRTMG in the rollout. Crucially, the use of a trustworthy physics-based radiation scheme ensures that the imposed radiative forcing is physically reliable.  Because the forcing is made consistently with the evolving state vector during autoregressive simulation, it enables a direct assessment of ACE’s hydrological response.  As ML emulation continues to mature, modular frameworks that allow individual components of the forecast system to be replaced and tested will become increasingly important \cite{Davenport2026}. 

The success of our methodology is not intrinsically limited to the E3SM model. The underlying architecture of ACE has proven flexible, having been successfully trained to emulate a variety of other Earth system models and reanalysis products, such as NOAA's fv3GFS \cite{Watt-meyer2025, Duncan2024}, and ML emulators also exist for the Community Atmosphere Model \cite{Chapman2025}. This demonstrated generality suggests that our approach could be applied across the international modeling community to rapidly assess fast atmospheric responses in diverse models. Furthermore, the profound computational efficiency of such emulators unlocks the ability to generate ensembles of unprecedented size \citep{paper_174}. This offers a powerful new avenue for robustly sampling internal atmospheric variability and disentangling it from the forced climate response.


While this study successfully addresses the fast atmospheric \cotwo\ response, a significant frontier is the characterization of slow climate responses. Here, we test the model’s response to clear-sky radiative perturbations. An important next step for ML emulators is to explicitly represent clouds, including cloud condensate and temperature-mediated cloud adjustments. Learning cloud adjustments to changes in radiative forcing agents has long been a central challenge for physically based climate models, and addressing it in ML emulators would allow cloud feedbacks to remain accountable to physical models. The broad objective is to extend this emulation framework to atmosphere–ocean coupled simulations, for instance by integrating ACE with three-dimensional ocean models \citep{Duncan2025, Dheeshjith2025} or slab ocean models \citep{Clark2025}. More generally, this work represents a step toward fully featured climate emulators that incorporate the key internal feedbacks and controllable forcings, such as aerosol emissions, land-surface change, and multiple categories of greenhouse gases, required for operational climate projection. The ultimate test for the new generation of ML emulators is their ability to reliably predict climate change.

\newpage
\section{Methods}\label{sec:methods}

We compare the responses of a historically trained ML emulator to heating from elevated \cotwo\  concentrations against the comparable responses from a full physics earth system model. We use two primary modeling frameworks: a comprehensive atmospheric general circulation model and a deep learning-based climate emulator.

\subsection{Physics-based Models}\label{ssec:EAMv2}

\subsubsection*{Model Description}

 The reference simulations were generated using EAMv2, the atmospheric component of the Energy Exascale Earth System Model (E3SM) \cite{Golaz2022}. Our simulations with EAMv2 are all branched from the 20230614.v2.LR.F2010 control run of E3SM used to train ACE.  We branched from years 64--73 withheld from the training data.  The control is based on the ``F2010'' (2010 CE radiative forcing) component set using the E3SMv2.LR configuration of E3SM.  The atmospheric dynamics are computed using a spectral element dynamical core on a cubed sphere with 30 spectral elements per cubed-sphere face.  This corresponds to a resolution of approximately 1${}^\circ$ in latitude and longitude.  

\subsubsection*{EAMv2 Reference Simulations with Different \cotwo\  Concentrations}

As a benchmark to assess the fast responses in ACE, we calculate the fast responses to different \cotwo\ levels in EAMv2.  We employ a version of EAMv2 without an active carbon cycle so that the concentration of \cotwo\  is a specified time-invariant constant throughout the atmosphere. 

Perturbations to all-sky radiative heating rates come from the difference in clear-sky radiative heating rates between N$\times$ and 1$\times$\cotwo.  We use clear-sky rather than all-sky calculations since ACE simulates the effects of clouds implicitly and thus does not provide prognostic cloud condensate that would be required for interactive all-sky radiative feedback (Section~\ref{ssec:ChangingCO2}).  EAMv2's radiative transfer calculations are handled by the Rapid Radiative Transfer Method for GCMs (RRTMG) \cite{Clough2005}. We employ the capability of EAMv2 to perform multiple diagnostic radiative transfer calculations with the atmospheric state held fixed.  When the \cotwo\  scaling factor N is set to 1, the simulations from our implementation of EAMv2 are bit-for-bit identical with the standard release of EAMv2, as required.  The modifications to inject heating from \cotwo\  into the atmosphere preserve the conservation of energy and water established with EAMv2's standard fixers to within double-precision roundoff.  In order to keep the pressure integral of the all-sky heating rates equal to the all-sky flux divergence across the atmospheric column, an equality required for energy conservation, the all-sky TOA fluxes are augmented by the change in clear-sky flux divergences between the top of atmosphere (TOA) and surface.  

The reason for applying these augmentations at TOA only is to avoid directly perturbing the interactive land surface by changes in the radiative fluxes incident on it from modifying \cotwo\  concentrations in the atmosphere.  To leading order, this means the land surface is subject to the same upper boundary conditions it experienced during the creation of the historical climatology used to train ACE.  This detail is important to ensure that the lower-boundary condition is held as close as possible to its time-mean state during ACE training.  Likewise, the \cotwo\  communicated from the atmosphere to the surface is for 1$\times$\cotwo\ conditions to ensure that the stomata (and their effects on evapotranspiration) are held to the historical conditions used to train ACE.  Evapotranspiration is responsible for approximately 10\% of the Earth's atmospheric moisture, and hence variations in it will directly impact precipitation.  

The moisture variable in ACE represents specific total water, i.e., the sum of specific humidity as well as liquid and ice condensates.  In ACE-RRTMG, we use this specific total water field as the specific (vapor) humidity input to RRTMG.  To perform the analogous simulation with EAMv2, we therefore also use specific total water as the specific (vapor) humidity input to RRTMG during the EAMv2 integrations with different levels of \cotwo. Replacing water vapor with specific total water as the moisture variable results in a $\mathcal{O}(1\%)$ change.  

Using this configuration, we generate a 96-member ensemble of 1-month-long integrations for a range of \cotwo\  concentrations: 0.5×, 1×, 2×, 4×, and 8× the 2010 CE level.  Each ensemble member is initialized on the first of the month using 8 years of the EAMv2 dataset that were not used to train ACE. In total, these simulations span a ten-year period, providing a statistically robust dataset to characterize the fast atmospheric response to \cotwo\  perturbations.

\subsection{Machine Learning Emulator}\label{ssec:ACE}

The machine learning component of this work utilizes the Allen Institute Climate Emulator (ACE), a model known to produce stable, long-term climate emulations \cite{Watt-meyer2023,Watt-meyer2025}. ACE is based on NVIDIA's FourCastNet Spherical Fourier Neural Operator (SFNO) architecture \cite{Bonev2023}. ACE is trained on the EAMv2 training dataset introduced in \cite{Duncan2024}.  ACE emulates an EAMv2 simulation in a ``perpetual 2010'' configuration, with the simulation using the annual cycle of SSTs from 2010 CE.  ACE includes three types of variables: forcing, prognostic, and diagnostic. For convenience, we reproduce the tables from \cite{Duncan2024} which show the ACE variables (Table~\ref{table:variable_list}) and the ACE vertical levels (Table~\ref{table:hybrid_levels}).  Notably, while some new versions of ACE have a feature to include a \cotwo\  variable as input, none of the versions used in the manuscript have this input field.

\begin{table}[t]
\centering
\caption{List of variables used in ACE. Forcing variables are input-only, diagnostic variables are output-only, and prognostic variables are both input and output by ACE.}
\label{table:variable_list}
\begin{tabular}{lllll}
\toprule
Symbol & Description & Units & Time & Variable Type \\
\midrule
$T_k$ & Air temperature & K & Snapshot & Prognostic \\
$T_s$ & Skin temperature of land or sea ice & K & Snapshot & Prognostic \\
$q_k^{T}$ & Specific total water (vapor + condensates) & kg/kg & Snapshot & Prognostic \\
$U_k$ & Wind speed in eastward direction & m/s & Snapshot & Prognostic \\
$V_k$ & Wind speed in northward direction & m/s & Snapshot & Prognostic \\
$p_s$ & Atmospheric pressure at surface & Pa & Snapshot & Prognostic \\

$RSW$ & Upward shortwave radiative flux at TOA & W/m$^2$ & Mean & Diagnostic \\
$OLR$ & Upward longwave radiative flux at TOA & W/m$^2$ & Mean & Diagnostic \\
$USW_{\mathrm{sfc}}$ & Upward shortwave radiative flux at surface & W/m$^2$ & Mean & Diagnostic \\
$ULW_{\mathrm{sfc}}$ & Upward longwave radiative flux at surface & W/m$^2$ & Mean & Diagnostic \\
$DSW_{\mathrm{sfc}}$ & Downward shortwave radiative flux at surface & W/m$^2$ & Mean & Diagnostic \\
$DLW_{\mathrm{sfc}}$ & Downward longwave radiative flux at surface & W/m$^2$ & Mean & Diagnostic \\
$LHF$ & Surface latent heat flux & W/m$^2$ & Mean & Diagnostic \\
$SHF$ & Surface sensible heat flux & W/m$^2$ & Mean & Diagnostic \\
$P$ & Surface precipitation rate (all phases) & kg/m$^2$/s & Mean & Diagnostic \\
$\left.\dfrac{\partial TWP}{\partial t}\right|_{\mathrm{adv}}$ 
& Tendency of total water path from advection 
& kg/m$^2$/s & Mean & Diagnostic \\
$S$ & Solar insolation & W/m$^2$ & Snapshot & Forcing \\
$T_{s,\mathrm{sea}}$ & Skin temperature over sea & K & Snapshot & Forcing \\
$\Phi_s$ & Surface geopotential & m$^2$/s$^2$ & Invariant & Forcing \\
$f_{\mathrm{land}}$ & Land grid cell fraction & N/A & Invariant & Forcing \\
$f_{\mathrm{ocean}}$ & Ocean grid cell fraction & N/A & Invariant & Forcing \\
$f_{\mathrm{ice}}$ & Sea-ice grid cell fraction & N/A & Invariant & Forcing \\

\bottomrule
\end{tabular}
\end{table}

\begin{table}[t]
\centering
\caption{Hybrid vertical coordinate parameters and reference pressures for ACE model level interfaces \cite{Duncan2024}. $I_k$ indexes EAMv2 model interfaces.  Pressure $p=a_k+b_k\,p_s$ where $p_s$ is the surface pressure.  $p_k^{ref}$ is computed for $p_s=1000$\ hPa.}
\label{table:hybrid_levels}
\begin{tabular}{ccccc}
\toprule
$k$ & $a_k$ [Pa] & $b_k$ [unitless] & $I_k$ & $p_k^{\mathrm{ref}}$ [hPa] \\
\midrule
0 & 10.0      & 0.0     & 0  & 0.1   \\
1 & 4943.694  & 0.0     & 19 & 49.4  \\
2 & 13913.118 & 0.0     & 30 & 139   \\
3 & 16254.503 & 0.10464 & 38 & 267   \\
4 & 12435.282 & 0.31152 & 44 & 436   \\
5 & 8945.939  & 0.50053 & 48 & 590   \\
6 & 5115.018  & 0.70804 & 53 & 759   \\
7 & 2027.536  & 0.87529 & 61 & 896   \\
8 & 0.0       & 1.0     & 72 & 1000  \\
\bottomrule
\end{tabular}
\end{table}

\subsection{Modifying the ACE Rollout to Calculate the Fast Response to Elevated Greenhouse Gases}\label{sec:methods_modified_ace_rollout}

A central innovation of our work is the coupling of the RRTMG radiation code to the ACE emulator, enabling it to simulate the response to \cotwo\  concentrations even though it was not trained on EAMv2 simulations with that scenario. We utilize an open-source version of RRTMG made available through the Python-based climt library \cite{Monteiro2018}. At present, radiation remains a computational bottleneck.  Each call to RRTMG requires approximately 8 seconds on a CPU, while each timestep with the columnwise ACE takes 0.35 seconds on an NVIDIA A100 GPU. Further optimization, including the adoption of GPU-accelerated radiation schemes, offers a clear path to substantially reducing this cost \cite{Pincus2019}.

During each inference time step, the coupled RRTMG module calculates the radiative forcing perturbation corresponding to a given \cotwo\  concentration ($N\times$\cotwo) relative to the $1\times$\cotwo\ state on which ACE was trained. We add the resulting perturbation to the temperature before advancing to the next time step (Figure~\ref{fig:rollout_modification}). This method allows the emulator to simulate the fast atmospheric adjustments to radiative forcing interactively during the rollout.  A critical feature of our experimental design is that ACE itself is passed no information that \cotwo\  has been perturbed from its recent historical concentration.  The only information passed to ACE are small perturbations to the atmospheric temperatures that fall well within the range of perturbations experienced during training against historical EAMv2 simulations.

When we apply the heating rate perturbations from RRTMG, we distinguish between the thermodynamic regimes of the troposphere and the stratosphere. For the troposphere, the perturbation from elevated \cotwo\  is calculated as an instantaneous change to atmospheric heating rates assuming  fixed background states. This assumption is valid for the troposphere with a 1-month simulation, as its thermal evolution is strongly coupled to the immense heat capacity of the ocean, which responds slowly to alterations in \cotwo\ (Figure~\ref{fig:cmip_ensemble}). However, this approach is inappropriate for the stratosphere, which is in radiative equilibrium and adjusts to perturbations on a much faster timescale of months independently of the ocean \citep{IPCC_2001_WGI_Ch_6}.  In the stratosphere, we must apply both forcing and stratospheric adjustment because the stratospheric state is evolving quickly in response to the changes to \cotwo.  Therefore, we model the stratospheric adjustment to \cotwo\ with an exponential damping term incorporating a timescale of 30 days for the lower stratosphere and 7.5 days for the upper stratosphere.  These values are chosen to represent the slower response time of the lower stratosphere as informed by empirical estimates from sudden stratospheric warming events \citep{Bloxam2021}.  Our perturbations take the following form:

\begin{equation}
\Delta Q(t) =
\begin{cases}
\left[-Q_{1 \times \text{CO}_2}^\text{net}+ Q_{N \times \text{CO}_2}^\text{net}\right] \delta t, & l > 1, \\
\left[-Q_{1 \times \text{CO}_2}^\text{net} + Q_{N \times \text{CO}_2}^\text{net}\right] \delta t \, e^{-t/\tau_l}, & l \leq  1 
\end{cases}
 \label{eqn:perturbation_rrtmg_form}
\end{equation}

\noindent where $l$ is the index of the model level, whereby $l=0$ refers to the topmost level and $l=7$ refers to the level nearest to the surface.  
The term $t$ denotes the number of hours since the initialization of the run and $\tau_l$ denotes the characteristic timescales for the upper and lower stratospheric responses.   
Equation~\ref{eqn:perturbation_rrtmg_form} shows that the temperature heating rate perturbations from RRTMG in the stratosphere are multiplied by an exponential decay term, while the troposphere is subject to the heating rate perturbations directly from RRTMG.

To calculate the fast precipitation response to \cotwo, we generate a large ensemble of short, 1-month-long simulations with both the ML emulator and the physics-based model. Ensemble members are initialized at the start of each month for 8 years, resulting in 96 ensemble members paralleling the approach employed in \cite{Stjern2023}. This approach contrasts with other fixed-SST experiments, such as the Precipitation Driver and Model Response Model Intercomparison Project \citep{Samset2016}, which often use a single, long, free-running simulation. We opt for an ensemble of short simulations for two reasons. First, it allows us to directly assess the statistically significant response within the first 7 days, the characteristic turnover timescale of water vapor in the atmosphere. Unlike one decadal simulation, an ensemble of short simulations effectively separates the forced climate response from chaotic weather noise  (Figure~\ref{fig:ensemble_spread_demo}). Second, the experimental design is computationally efficient.  Each ensemble member can be run in parallel, allowing us to leverage the computational speedup of ACE more effectively than with a single serial integration.  Each ensemble member only requires a single GPU, so if more than one GPU is available at one time, an ensemble of short simulations will have a faster time to solution than a single long simulation.  At each six-hourly inference timestep, the 1 $\times$ CO$_2$ and $N \times$ CO$_2$ calls to RRTMG take 8 seconds each on an AMD EPYC 7763 CPU, and the modified columnwise ACE step takes approximately 0.35 seconds on an NVIDIA A100 GPU.

\subsection{Criteria for a Successful Emulator of Fast Responses}\label{ssec:criteria_for_success}

We enumerate the following criteria for a successful emulator of fast responses:

\begin{enumerate}
    \item Correct representation of the globally averaged and spatially resolved fast responses in the hydrological cycle to \cotwo\ perturbations
    \item Correct response to the idealized perturbation (described in Section~\ref{ssec:idealized_test})
    \item Low RMSE of predicted fields on 10-year rollouts in the base climate
    \item Small moisture conservation errors
\end{enumerate}

The default ACE requires enhancements to pass \#1 (Figure~\ref{fig:global_mean_response_defaultace} versus Figure~\ref{fig:global_mean_response} and Figure~\ref{fig:spatial_patterns_defaultace} versus Figure~\ref{fig:spatial_patterns}) and \#2 (Figure~\ref{fig:idealized_test_ace}).  Our proposed column-local architecture passes \#1-4.  

For the fast response experiment in this manuscript, a 10-year rollout is unnecessary; the fast responses are measured in an ensemble of short, 1-month rollouts.  However, since these emulators are designed for climate applications, we seek to preserve stability on long timescales (\~10 years) as we design modified ACE architectures. Therefore, we include criterion \#3 in our determination of a successful emulator.

\subsection{Evaluating ACE Fast Responses with an Idealized Profile}~\label{ssec:idealized_test}

We also test ACE by applying an idealized heating rate perturbation.  This test is designed to mimic the dipole vertical profile of the net radiative effects of \cotwo\ (Figure~\ref{fig:heating_rate_perturbations}) that combines heating in the troposphere with cooling in the stratosphere.  We adopt an idealized perturbation of the net radiative heating rates that mimics the dipole heating by \cotwo\ using the following structure:

\begin{equation}
\delta Q^{\text{net}}_{\text{idealized}} =
\begin{cases}
-h, & l \leq l_{\text{strat}}, \\
h, & l > l_{\text{strat}}.
\end{cases}
\label{eqn:perturbation_idealized}
\end{equation}

\noindent where $h$ is an imposed heating rate perturbation in units of \(\mathrm{K\,day^{-1}}\), $l \ge 0$ indexes model levels in order of increasing pressure, and $l_{\text{strat}}$ is the highest value of $l$ for layers in the stratosphere. Since ACE allocates its top two model levels to the stratosphere and the bottom six to the troposphere (Table~\ref{table:hybrid_levels}), $l_{\text{strat}} = 1$ for our specific application.  However, the specific value of $l_{\text{strat}}$ applicable to ACE is not an essential component of the test and could be readily changed for emulators with other vertical discretizations of the atmosphere.

\subsubsection*{Column-integrated energetic scaling}

We estimate the expected precipitation response to a prescribed atmospheric heating perturbation using a column-integrated energy balance.


Let $p$ denote atmospheric pressure in units of hPa, $p_{\text{strat}} = 139$~hPa represent the pressure at the tropopause (Table~\ref{table:hybrid_levels}), and $p_s = 985$~hPa represent the global-mean surface pressure.  In our idealized dipole, stratospheric cooling ($p \le p_{\text{strat}}$) approximately cancels the heating in the upper troposphere ($p_{\text{strat}} < p \le 2\,p_{\text{strat}}$). The net column heating therefore applies to the remainder of the troposphere:
\begin{equation}
\Delta p = p_s - 2\,p_{\text{strat}}.
\end{equation}
Again, the specific value of $p_{\text{strat}}$ applicable to ACE is not an essential component of the test and could be readily changed for emulators with other vertical discretizations of the atmosphere.

Assuming hydrostatic balance and a constant gravitational acceleration of 
\(
g \approx 9.81~\mathrm{m\,s^{-2}},
\)
the column mass of the perturbed layer is
\begin{equation}
m = \frac{\Delta p}{g}
= 7.21 \times 10^3\ \mathrm{kg\,m^{-2}}.
\end{equation}


\noindent Using a constant specific heat capacity of dry air at constant pressure,
\begin{equation}
c_p = 1005~\mathrm{J\,kg^{-1}\,K^{-1}},
\end{equation}
the column-integrated heat capacity of the perturbed atmospheric column is
\begin{equation}
C = c_p\,m  = c_p\,\frac{\Delta p}{g}= 7.25 \times 10^6~\mathrm{J\,m^{-2}\,K^{-1}}.
\end{equation}


\noindent Given a value of $h$ in units of K/day from Equation~\ref{eqn:perturbation_idealized}, the corresponding column-integrated energy imbalance is
\begin{equation}
\Delta R = \frac{C}{L_{\text{day}}}\,h = \frac{c_p}{L_{\text{day}}}\,\frac{\Delta\,p}{g}\,h
\approx 83.9\,h~\mathrm{W\,m^{-2}}. \label{eq:EnergyImbalance}
\end{equation}
\noindent where the length of day $L_{\text{day}} = 86400$ seconds/day.
\subsubsection*{Connections between atmospheric energetic inputs and precipitation}

Assuming that the atmosphere maintains thermal equilibrium, and assuming that the upper boundary conditions on the atmosphere are fixed, this energy imbalance must be balanced by a combination of energy inputs from the surface and/or by internal work done by the atmospheric gases.  That work could take the form of either accelerating or decelerating the large-scale atmospheric flow or altering the gravitational potential energy of the atmosphere through alterations to its vertical structure.  Since there is no evidence in our EAMv2 simulations that either form of work is occurring, the column energy imbalance given by Equation~\ref{eq:EnergyImbalance} must be balanced by changes in the fluxes of energy from the Earth's surface to the atmosphere.  The four relevant fluxes are the longwave, shortwave, latent, and sensible heat fluxes.  In our experiments where surface temperatures and radiative properties are held fixed, the upward shortwave and  longwave fluxes from the surface to the atmosphere are therefore also held fixed and hence cannot act to restore atmospheric energy balance.  This implies that perturbations to the latent and sensible heat fluxes act to restore energy balance, and this mechanism is observed to be operative in our EAMv2 simulations.  

The fraction of the energy input  by latent heating is given 
\begin{equation}
f_B = \frac{LH}{LH + SH} = \frac{1}{1+B},
\end{equation}
where \(LH\) and \(SH\) denote latent and sensible heat fluxes, respectively, and \(B\) is the Bowen ratio \(SH/LH\). Assuming that \(f_B\) remains unchanged under the perturbation, only a fraction \(f_B\) of the energy imbalance induced by the perturbation $h$ is balanced by latent heating with the remainder balanced by sensible heating. In the EAMv2 1$\times$\cotwo\ control, the globally averaged time-mean surface fluxes are \(SH = 20\)~W/m${}^2$ and \(LH = 87\)~W/m${}^2$, which combine to yield \(B = 0.23\) and \(f_B = 0.81\). Since ACE is trained on output from EAMv2, we hypothesize that this relative partitioning between latent and sensible heat flux pertains to ACE and our modifications to it as well. We have verified that $B$ is unchanged in the N$\times$\cotwo\ experiments using EAMv2 relative to its 1$\times$\cotwo\ control and that the Bowen ratios for EAMv2 and columnwise ACE agree to within 2.5\%. Note that if this diagnostic is applied to an emulator trained on a different climatological data set, then the value of $f_B$ should be recomputed for this climatology.

If we treat RCE as a closed hydrological cycle in quasi equilibrium, then the global-mean sources and sinks of water must balance and be nearly identical.  In the troposphere where RCE is operative, essentially all the water vapor is due to evaporation and evapotranspiration from the surface.  This implies that the global mean precipitation and latent heat flux should be nearly identical when expressed in common units, and that this approximate equality should also apply to perturbations to RCE.   This balance is manifest in our simulations with EAMv2 (Figure~\ref{fig:global_mean_response}). These connections across the climate system imply that a fraction $f_B$ of the column energy imbalance $\Delta R$ (Equation~\ref{eq:EnergyImbalance}) should be rebalanced by the latent heating from the formation of liquid and solid precipitation.

\subsubsection*{Conversion from energy flux to precipitation}

Based on these connections, the global water balance, together with the fraction $f_B$ of the column energy imbalance mediated by latent heating associated with condensation and fusion, imply that the precipitation should change by:
\begin{equation}
\Delta P = f_B\,\frac{\Delta R}{L_v + f\,L_f},
\label{eq:PfromDeltaR}
\end{equation}
\noindent where the latent heats of condensation and fusion are
\begin{eqnarray}
L_v &=& 2.257 \times 10^6~\mathrm{J\,kg^{-1}} \\
L_f &=& 0.334 \times 10^6~\mathrm{J\,kg^{-1}}
\end{eqnarray}
\noindent and where $f$ with $0\le f \le 1$ is the fraction of water vapor that first condenses and then fuses to form ice and snow.

Numerical evaluation of Equation~\ref{eq:PfromDeltaR} for ACE yields
\begin{eqnarray}
\Delta P  
&=& \frac{h}{1+ 0.148\,f}\, 3.01 \times 10^{-5}\ \mathrm{kg\,m^{-2}\,s^{-1}} \\
&=& \frac{h}{1+ 0.148\,f}\, 2.60\,\mathrm{mm\,day^{-1}} 
\end{eqnarray}
This derivation shows that our idealized heating profile should induce a response in precipitation that is bounded between 
\begin{equation}
    \Delta P = \left\lbrace \begin{array}{ll} 2.27\,\mathrm{mm\,day^{-1}} & f = 1 \\
    2.60\,\mathrm{mm\,day^{-1}} & f = 0
    \end{array} \right.
\end{equation}
These limits are plotted as dashed lines on Figure~\ref{fig:idealized_test_ace}.
In our idealized test, we check whether the ML emulator response falls within this range.  Emulators that fall outside this range are  violating either the conservation of energy or water, or both, and are therefore manifestly unsuitable for emulating perturbations to RCE.  As the figure shows, the default ACE fails this idealized test, while our modified ACE passes it handily.

This test can be readily applied to other ML weather and climate emulators as a convenient way to assess the fast responses without the need for coupling to physics-based radiative transfer codes.

\subsection{Modifying the ACE Architecture}\label{sec:ace_modifications}

\subsubsection{Motivation for Changing the ACE Architecture}

As shown in Figures~\ref{fig:global_mean_response_defaultace} and~\ref{fig:spatial_patterns_defaultace}, the default ACE architecture in \cite{Duncan2024} does not correctly reproduce the response of the hydrological cycle to changes in \cotwo\ simulated by EAMv2 (Section~\ref{ssec:criteria_for_success}). The default architecture infers that precipitation will increase rather than decrease monotonically in response to increasing \cotwo.  Hence both the sign and tendency of the precipitation response differ relative to EAMv2.  In addition, the latent heat flux decreases in response to elevated \cotwo.  Since the latent heat flux supplies roughly 90\% of the atmospheric moisture that condenses to form precipitation, the opposite-signed tendencies in these quantities imply that default ACE does not conserve moisture in response to perturbations in \cotwo. 

The default ACE's spatial pattern in response to elevated \cotwo\  (Figure~\ref{fig:spatial_patterns_defaultace}) shows an increase in precipitation throughout most of the subtropics, midlatitudes, and poles.  We speculate this occurs due to a spurious spatial correlation learned during training. With elevated \cotwo, tropospheric temperatures increase. Since the tropics are warmer and rainier than the rest of the world, the emulator associates warmer air with more rain.  To make ACE robust to the \cotwo\  perturbations, we modify the architecture while using the same training dataset.  These modifications are enabled by the flexible ACE platform (https://github.com/ai2cm/ace) and its seamless support for modular changes.

\subsubsection{Description of Modifications to the ACE Architecture}\label{ssec:ace_modification}

Our columnwise ACE architecture has three modifications from the default ACE \cite{Duncan2024}. First, a nearly column-local neural network predicts all diagnostic variables.  This network is ``nearly'' column-local because it predicts the diagnostic variables using a 3 grid cell $\times$ 3 grid cell input patch.  (A purely column-local architecture would make predictions only using the vertical profiles from 1 grid cell.)  In this network, we use discrete-continuous spherical convolutions \cite{Ocampo2022, Bonev2025} to preserve spherical geometry.  The diagnostic network forces the diagnostic variables (such as precipitation and latent heat flux) to be predicted only using the nearby vertical columns of the atmosphere \cite{You2024, chen2025data}.  By default, ACE predicts the diagnostic variables using SFNO, which utilizes nonlocal information for its predictions through a global spherical harmonic transform.  The diagnostic architecture is a four-layer neural network.  The first layer is a discrete-continuous convolutional layer with a 3$\times$3 kernel \cite{Bonev2023, Ocampo2022}, and the remaining layers are fully connected layers. After each intermediate layer, a rectified linear unit activation function is used. 

Second, in the columnwise ACE, the prognostic variables are predicted as sums of two neural networks: SFNO and a separate column-local neural network. The default ACE only uses SFNO to predict the prognostic variables.  This second modification is present in other versions of FourCastNet used in \cite{Bonev2023, paper_173}, and we add it to ACE.   The prognostic neural network is a one-layer fully connected neural network.

A diagram of the prognostic and diagnostic column-local networks is provided in Figure~\ref{fig:architecture_diagram}. In the columnwise ACE, there are three neural networks: SFNO, the columnwise network for diagnostic variables, and the columnwise network for prognostic variables.  Compared to existing work, a unique aspect of our approach is that all three of these neural networks are trained jointly.  This makes it possible to include moisture conservation constraints as in \cite{Watt-meyer2025}, even though some moisture variables (precipitation, evaporation, and moisture advection) are produced by the diagnostic network while others (specific total water at each model level) are produced by SFNO and the columnwise prognostic network.  Additionally, NeuralGCM shows that online training of multiple model components improves both stability and performance \citep{Kochkov2024}. In ACE, we find that joint optimization is essential for effective coupling between components.  When networks are trained separately, the model fails to reproduce the correct magnitude of the precipitation response. Existing work proposes adding diagnostic variables \citep{Mitra2023, Pathak2022} to pretrained ML emulators that have a fixed set of prognostic variables. For climate change tests, this setup may not lead to the correct climate response. 

Our third modification to the ACE architecture is the addition of constraints during training.  Dry air is conserved; moisture is conserved; and moisture, precipitation rate, and radiative fluxes are forced to be positive.  These constraints were introduced in ACE2 \cite{Watt-meyer2025}, which is a successor to the ACE E3SMv2 model architecture used in \cite{Duncan2024}.   We find that these changes are not the major changes necessary for the correct fast response to elevated \cotwo\ (Section~B.2.1), but we include them because it is intrinsically advantageous to have an ML climate emulator that conserves moisture and dry air.

In order to arrive at the above-chosen architecture, we conducted a non-exhaustive set of hyperparameter tuning experiments in which we assessed the effect of several architectural design choices. We describe the results of hyperparameter tuning experiments in the Supplemental Information.


\backmatter



\bmhead{Acknowledgements}

This research was supported by the Director, Office of Science, Office of Biological and Environmental Research of the U.S. Department of Energy under Contract No. DE-AC02-05CH11231 and by the Regional and Global Model Analysis Program area within the Earth and Environmental Systems Modeling Program. The research used resources of the National Energy Research Scientific Computing Center (NERSC), which is also supported by the Office of Science of the U.S. Department of Energy, under Contract No. DE-AC02-05CH11231. The computation for this paper was supported in part by the DOE Advanced Scientific Computing Research (ASCR) Leadership Computing Challenge (ALCC) 2023-2024 award ``Huge Ensembles of Weather Extremes using the Fourier Forecasting Neural Network'' to William Collins (LBNL) and the 2024-2025 award ``Huge Ensembles of Weather Extremes using the Fourier Forecasting Neural Network'' to William Collins (LBNL). This research was supported in part by Lilly Endowment, Inc., through its support for the Indiana University Pervasive Technology Institute. This research was also supported in part by the Environmental Resilience Institute, funded by Indiana University's Prepared for Environmental Change Grand Challenge initiative. We are grateful to our colleague Christopher Bretherton for helpful discussions and feedback on this work.

We acknowledge the World Climate Research Programme, which, through its Working Group on Coupled Modelling, coordinated and promoted CMIP6. We thank the climate modeling groups for producing and making available their model output, the Earth System Grid Federation (ESGF) for archiving the data and providing access, and the multiple funding agencies who support CMIP6 and ESGF.  We acknowledge access to observational, gridded, and reanalysis data in the CMIP6 Data Archive,   
supported by the Program for Climate Model Diagnosis and Intercomparison (PCMDI).  This data archive is funded by the Climate and Environmental Sciences Division of the DOE Office of Science and is performed under the auspices of the U.S. Department of Energy by Lawrence Livermore National Laboratory under contract DE-AC52-07NA27344. The E3SM model and simulation data were supported as part of the Energy Exascale Earth System Model (E3SM) project, funded by the U.S.
Department of Energy, Office of Science, Office of Biological and Environmental Research.

\subsection*{Competing Interests}

The authors declare no competing interests.

\subsection*{Author Contributions}

A.M. and W.D.C. conceptualized the study. All authors contributed to the methodology. Software was developed by A.M., W.D.C., T.A.O., P.B.G., S.Z., S.S., J.P.C.D., O.W.-M., B.B., and T.K. Formal analysis was performed by A.M., W.D.C., T.A.O., P.B.G., S.Z., S.S., J.P.C.D., and O.W.-M. Resources were provided by W.D.C., K.K., and M.S.P. All authors contributed to writing the original draft and to the review and editing of the manuscript. W.D.C. supervised the project. W.D.C. and T.A.O. acquired funding.

\noindent

\bigskip


\begin{appendices}

\newpage
\section{Extended Data Figures}\label{sec:extended_data_figures}

\begin{figure}[h]
    \centering
    \includegraphics[width=\linewidth]{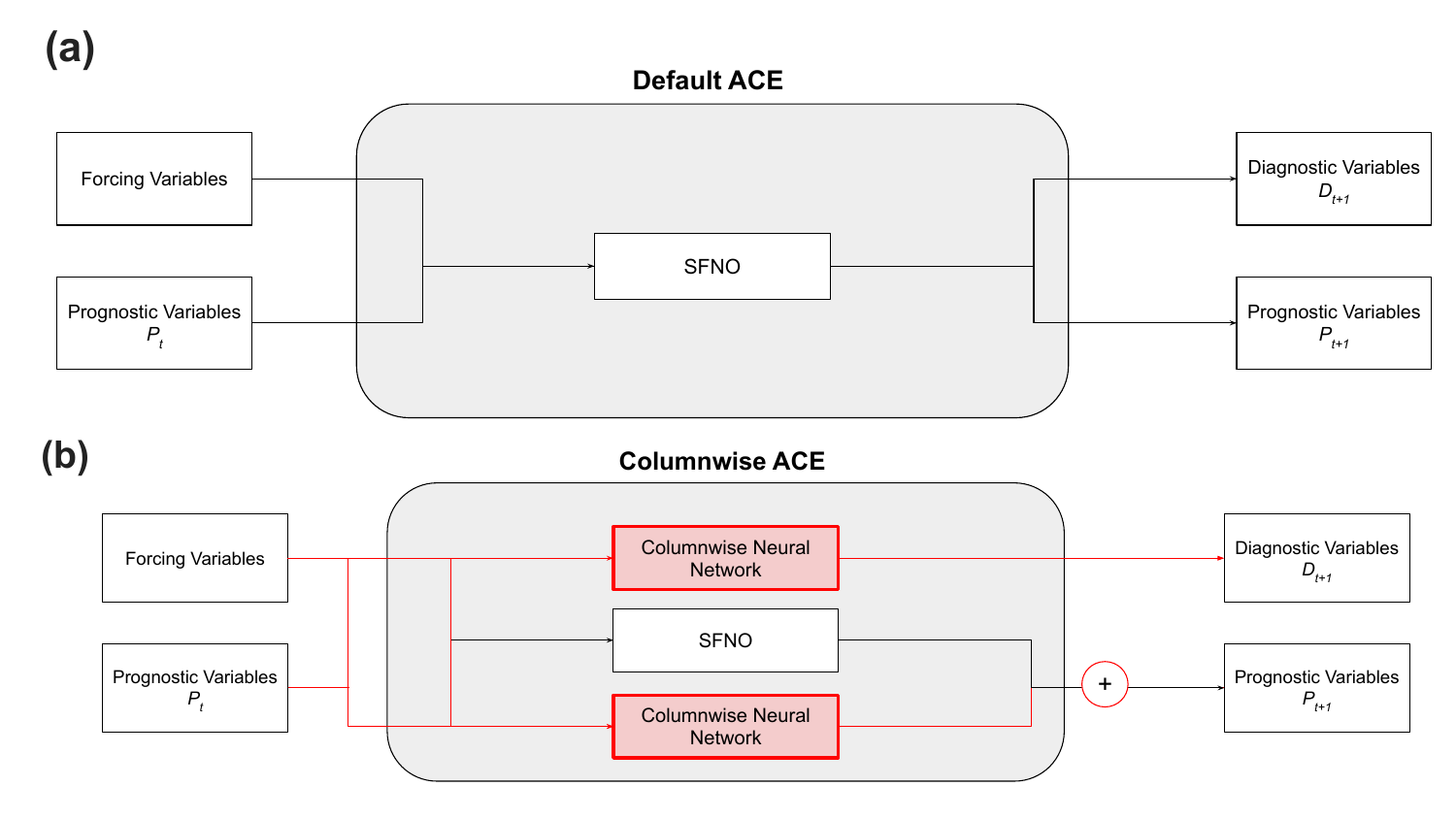}
    \caption{\textbf{Diagram of ACE Architecture Changes for Columnwise Diagnostics}. ACE has input-only variables (called forcing variables), output-only variables (called diagnostic variables), and variables that are both input and output (called prognostic variables).  Panel~(a): the default ACE architecture uses the Spherical Fourier Neural Operator (SFNO) architecture for all variables \cite{Duncan2024}.  SFNO is not a column-local architecture; rather, it is a global architecture that makes predictions using nonlocal information via a spherical harmonic transform.  Panel~(b): our proposed ACE architecture has two modifications depicted in red relative to the default ACE architecture.  First, the diagnostic variables are diagnosed solely based on a column-local neural network architecture.  Second, the prognostic variables are calculated from the sum of two networks: the SFNO and another column-local architecture.}
    \label{fig:architecture_diagram}
\end{figure}

\begin{figure}
    \centering
    \includegraphics[width=\linewidth]{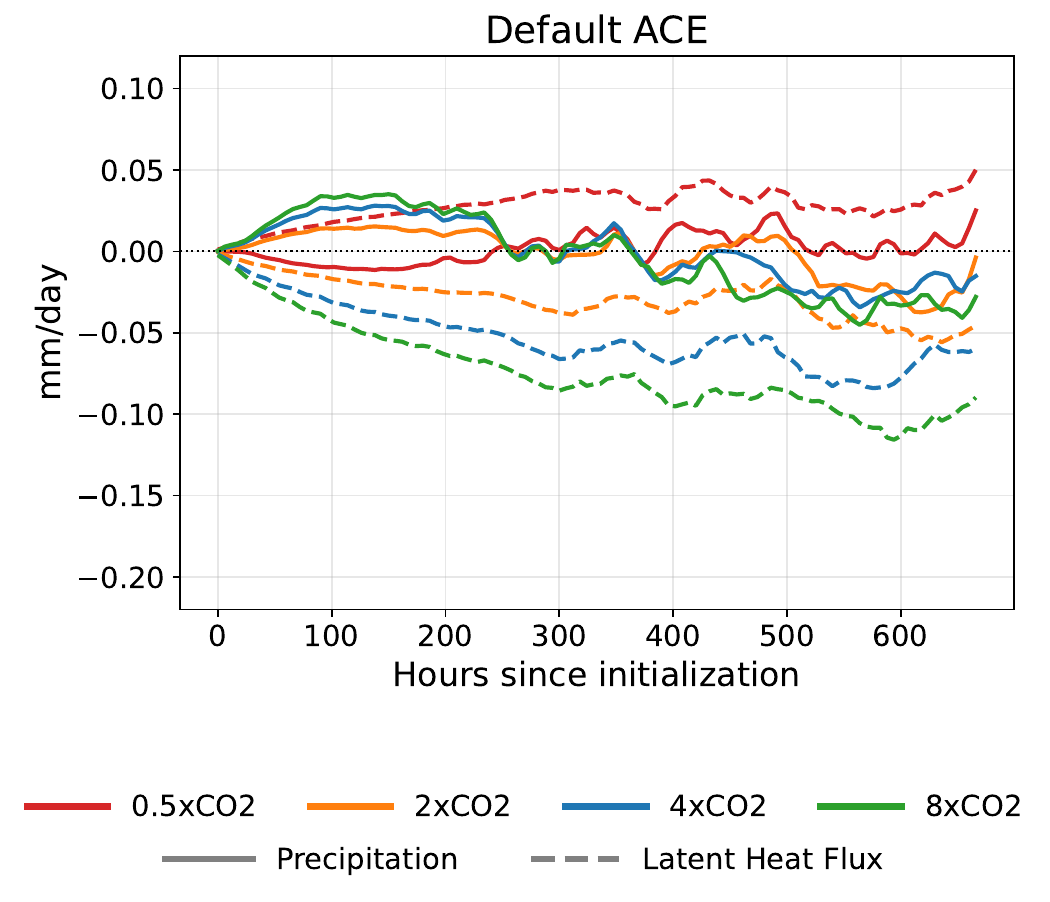}
    \caption{\textbf{Global Mean Fast Response to Instantaneous Changes in \cotwo\ 
Concentrations}. Same as Figure~\ref{fig:global_mean_response} but for the default ACE architecture  \cite{Duncan2024}. This architecture does not contain the columnwise modifications described in Section~\ref{sec:ace_modifications}.}
    \label{fig:global_mean_response_defaultace}
\end{figure}

\begin{figure}
    \centering
    \includegraphics[width=\linewidth]{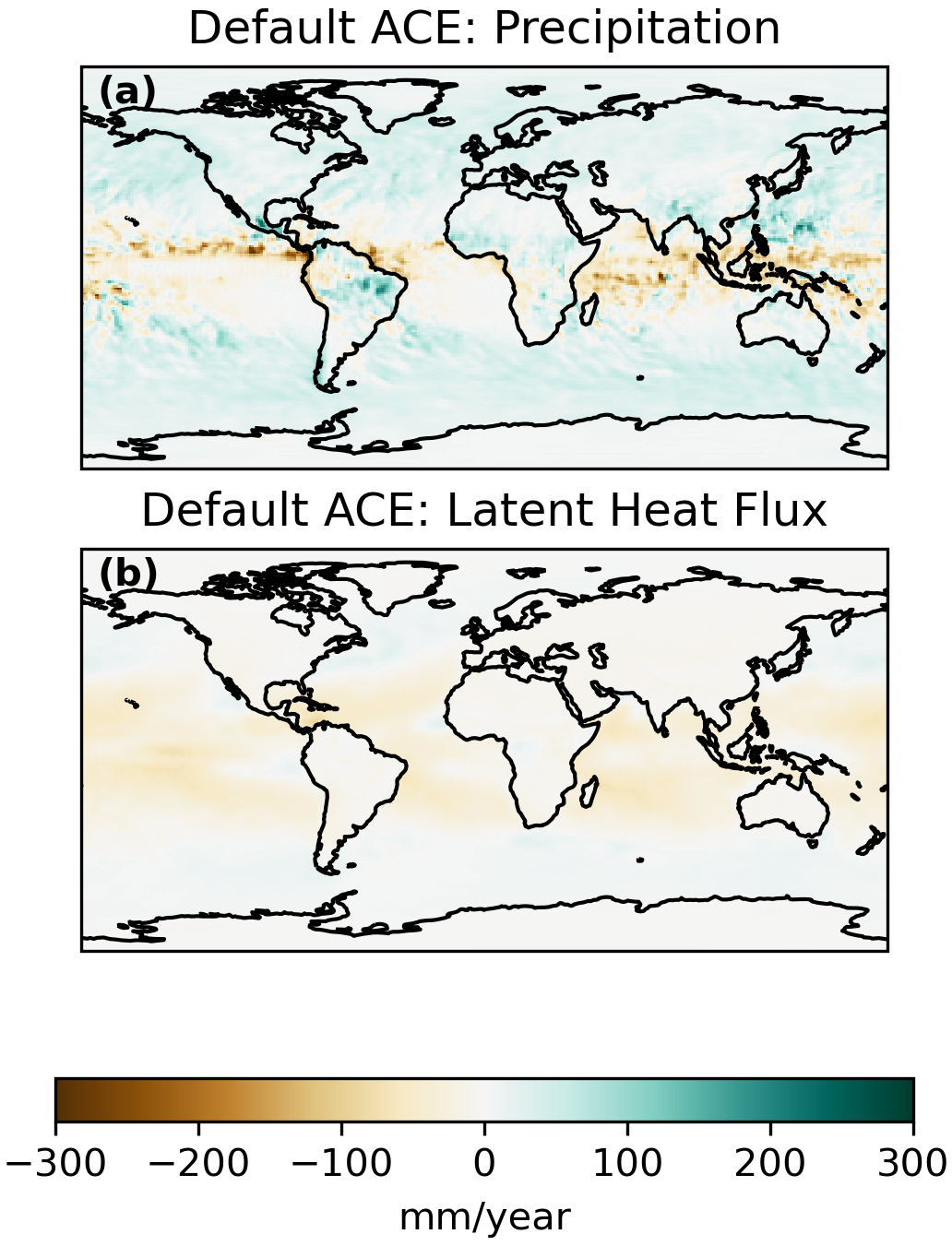}
    \caption{\textbf{Spatial Pattern of Fast Responses of Precipitation and Latent Heat Flux to 8$\times$\cotwo}. Same as Figures~\ref{fig:spatial_patterns}(a) and~(b) but for the default ACE architecture  \cite{Duncan2024}. This architecture does not contain the columnwise modifications described in Section~\ref{sec:ace_modifications}.}
    \label{fig:spatial_patterns_defaultace}
\end{figure}

\begin{figure}
    \centering
    \includegraphics[width=\linewidth]{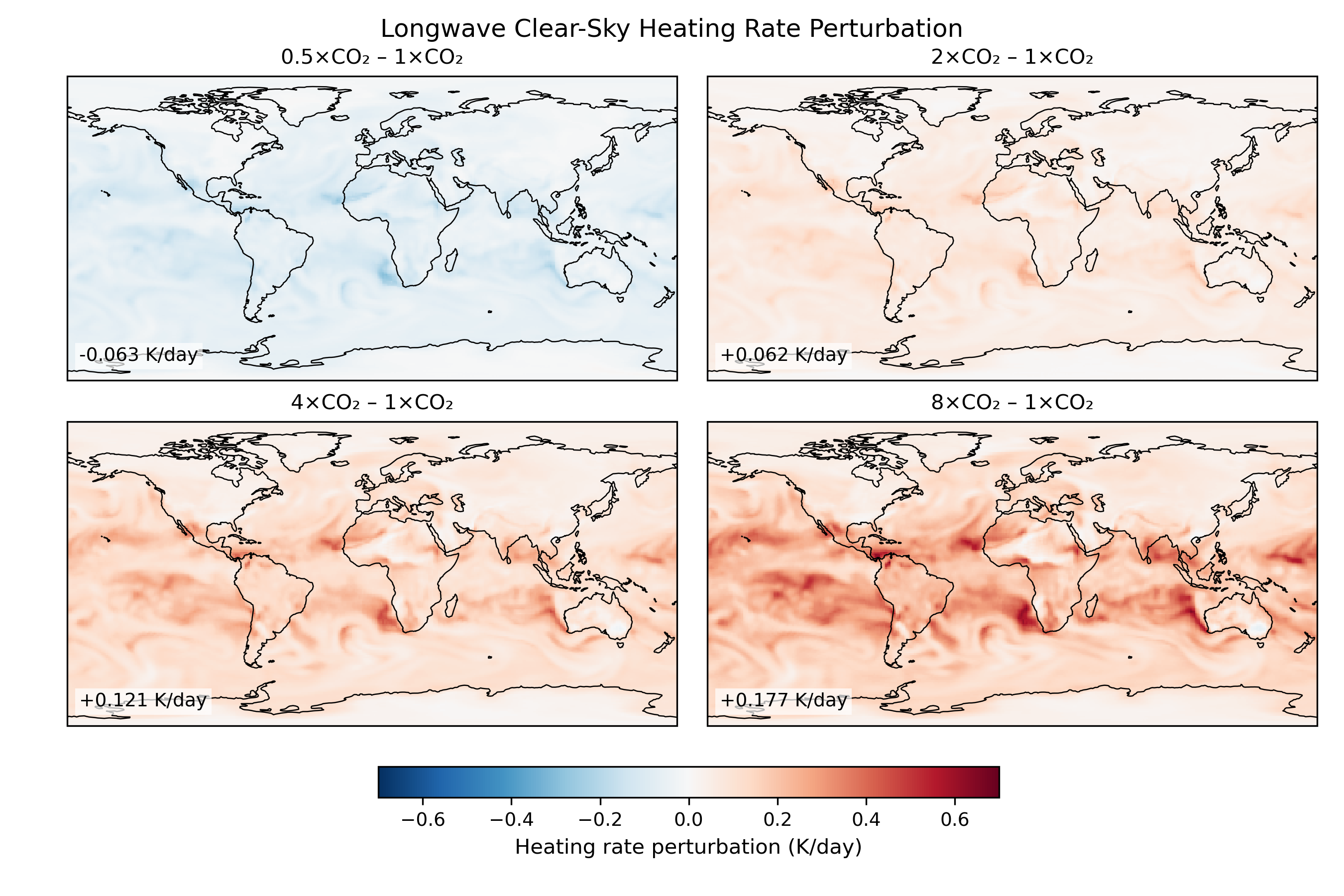}
    \caption{\textbf{Visualization of Heating Rate Perturbations}.  Longwave clear-sky heating rate perturbations for concentrations of \cotwo\ set to 0.5, 2, 4, and 8$\times$ its concentration in 2010 CE. Heating rate perturbations are calculated from RRTMG using Equation~\ref{eqn:perturbation_rrtmg_form}.  Heating rates are shown for a single representative timestep and apply to temperatures $T_7$ in layer~7 of ACE (Table~\ref{table:hybrid_levels}), i.e., the temperatures at the model level closest to the surface.  The global means are shown in the boxes in the bottom left of each panel.}
    \label{fig:heating_rate_perturbation_vis}
\end{figure}

\begin{figure}
    \centering
    \includegraphics[width=\linewidth]{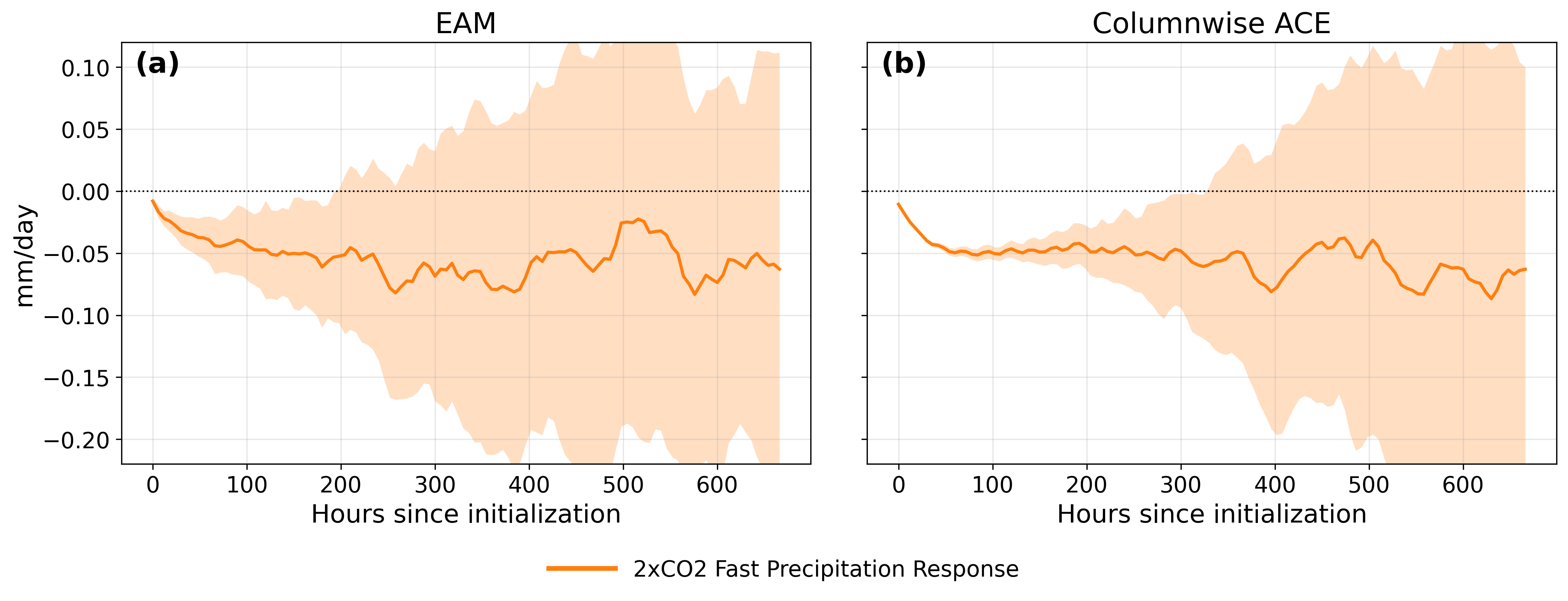}
    \caption{\textbf{Ensemble Spread as a Function of Lead Time}. Same as the 2 $\times$ \cotwo\ simulation shown in Figure~\ref{fig:global_mean_response}, but the shading denotes the standard deviation of the ensemble response as a function of the number of hours from initialization.  }
    \label{fig:ensemble_spread_demo}
\end{figure}

\begin{figure}
    \centering
    \includegraphics[width=1.1\linewidth]{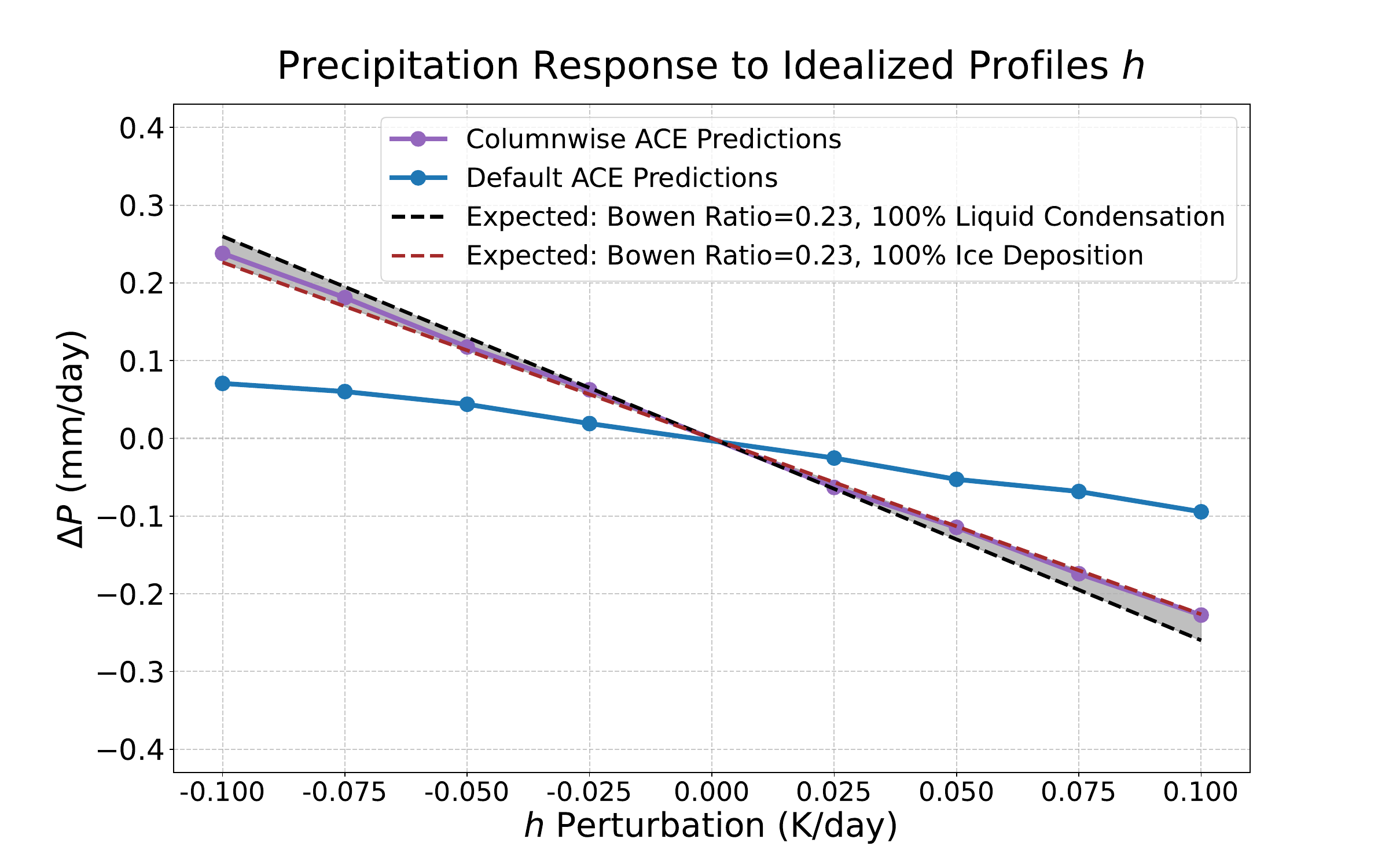}
    \caption{\textbf{ACE Response to Idealized Heating Rate Perturbations}.  The response to the idealized heating rate perturbation in Equation~\ref{eqn:perturbation_idealized} is shown for the default ACE and the columnwise ACE.  The expected response to the perturbation is calculated assuming a global-mean Bowen ratio of 0.23 obtained from EAMv2.  The shading denotes the expected range of the response, based on water vapor only condensing to liquid water (in which case $L_v$ is used) or exclusively depositing  as ice and snow (in which case $L_v+L_f$ is used).  See Section~\ref{ssec:idealized_test} for the derivations. }
    \label{fig:idealized_test_ace}
\end{figure}

\begin{figure}
    \centering
    \includegraphics[width=\linewidth]{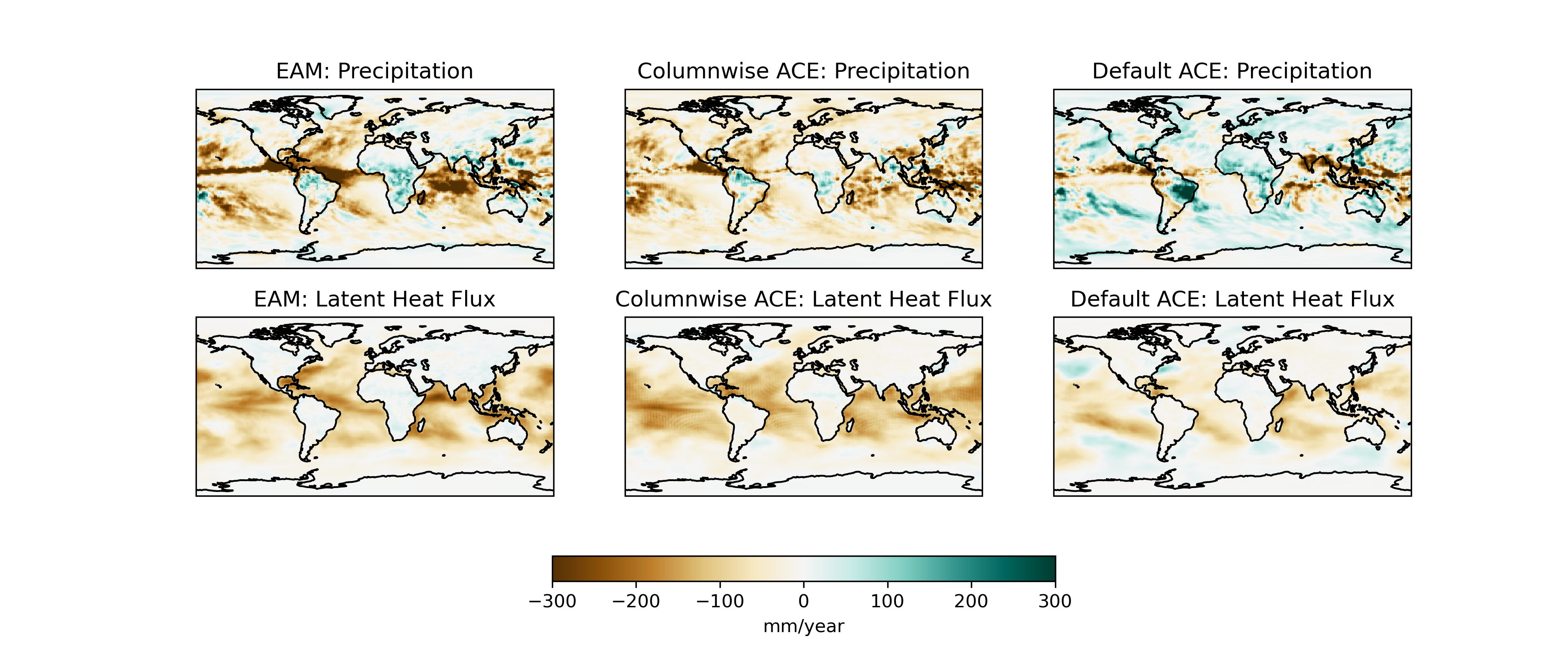}
    \caption{\textbf{Spatial Pattern of Precipitation and Latent Heat Flux Fast Response to 8$\times$\cotwo}. Same as Figures~\ref{fig:spatial_patterns} and~\ref{fig:spatial_patterns_defaultace} but the spatial patterns are averaged over 28 days of simulation.}
    \label{fig:spatial_patterns_alldays}
\end{figure}

\begin{figure}
    \centering
     \includegraphics[width=\linewidth]{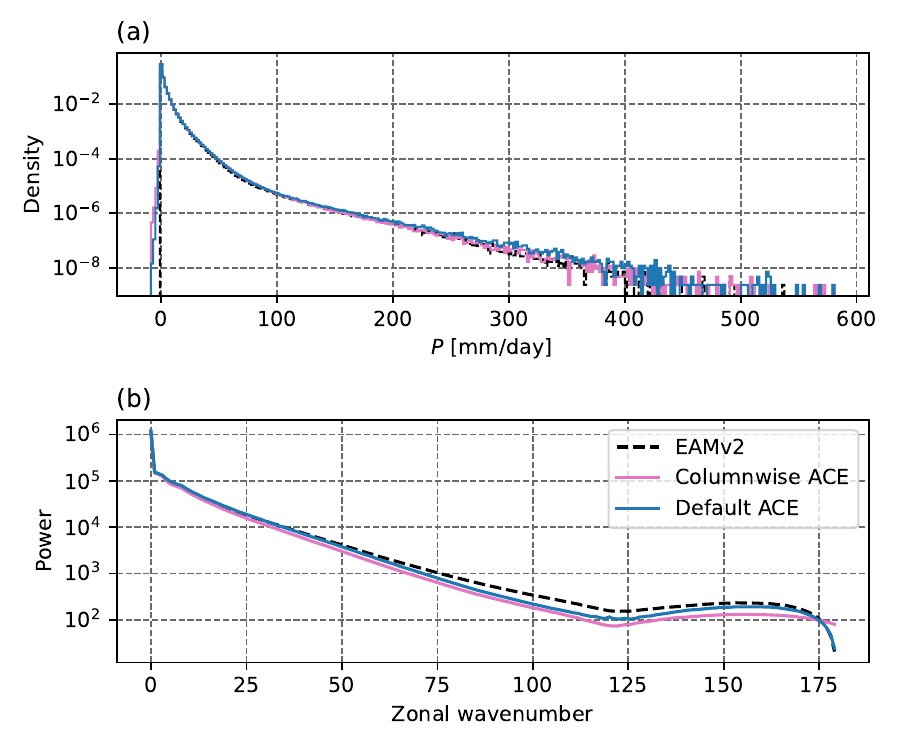}
    \caption{\textbf{Histogram and Zonal Spectra of Precipitation}.  Panel~(a): the histogram of daily mean precipitation for the default ACE, columnwise ACE, and EAMv2 training dataset.  Panel~(b): the meridional-mean zonal power spectrum of daily mean precipitation across all grid points over 10 years.}
    \label{fig:precip_hist_spectra}
\end{figure}

\begin{figure}
    \centering
    \includegraphics[width=0.5\textheight]{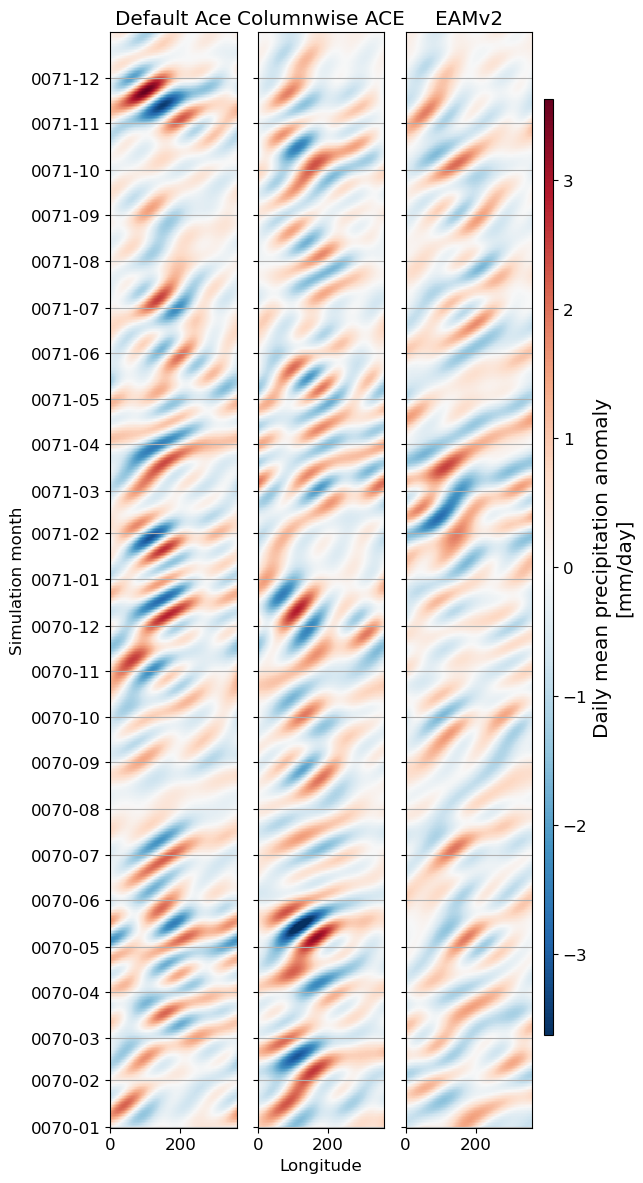}
    \caption{\textbf{ACE Precipitation Hovm\"oller Diagram}.  Hovm\"oller diagrams of $1 \times \text{CO}_2$ daily mean tropical-mean precipitation are shown over two years with a bandpass filter to retain 20–100 day periods. Diagrams use the default ACE, columnwise ACE, and EAMv2 training dataset in the left, center, and right columns, respectively. D}
    \label{fig:precip_hovmoller}
\end{figure}

\begin{figure}
    \centering
    \includegraphics[width=\linewidth]{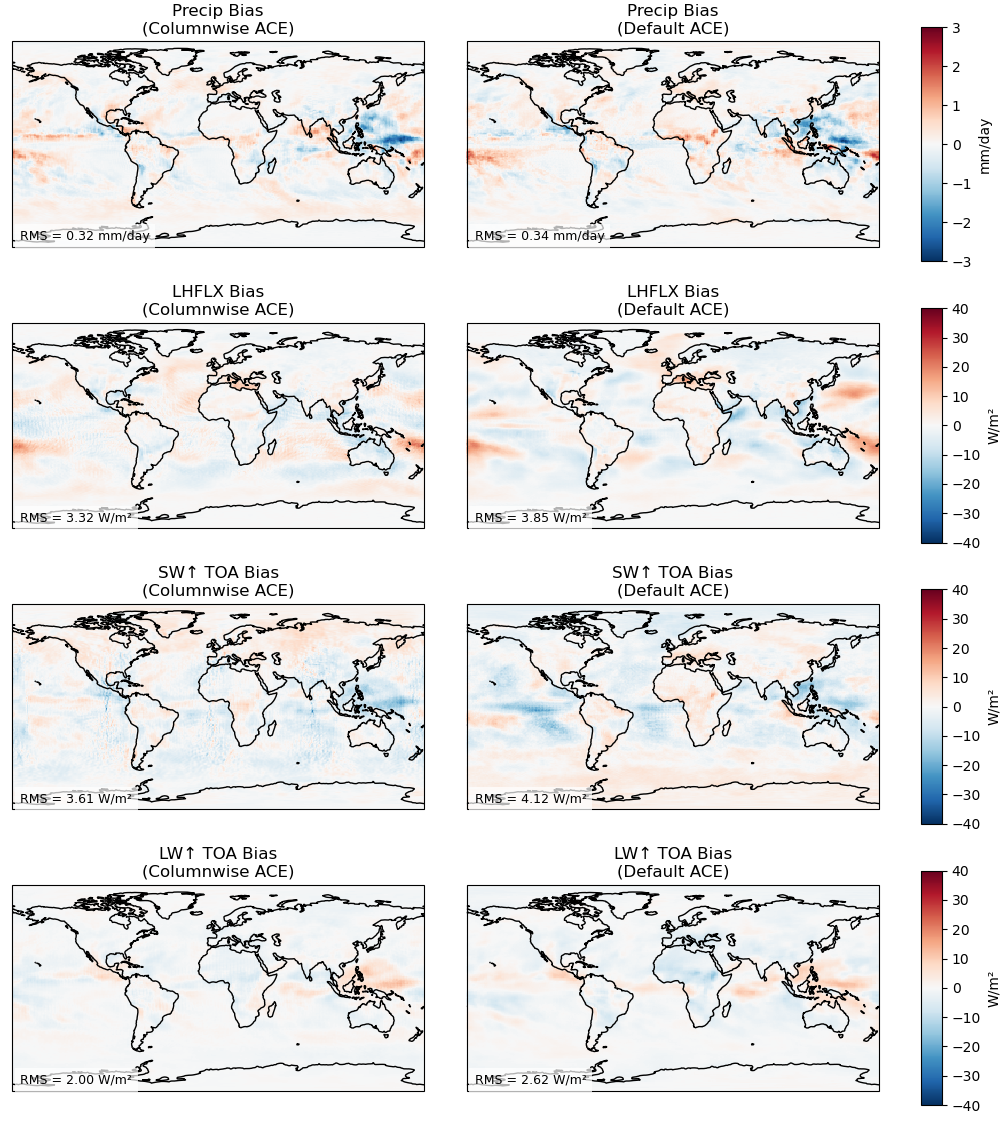}
    \caption{\textbf{Spatial Pattern of Bias}. Bias maps comparing the columnwise ACE (left column) to the default ACE (right column).  Bias is calculated by comparing an 8-year ACE rollout to the corresponding mean values of the EAMv2 out-of-sample test set.  Bias maps are shown for precipitation (``Precip"), latent heat flux (``LHFLX"), outgoing shortwave at the top of the atmosphere (``SW$\uparrow$ TOA"), and outgoing longwave at the top of the atmosphere (``LW$\uparrow$ TOA"). Values in the bottom left corner of each figure denote the global latitude-weighted root mean squared bias (RMSB).}
    \label{fig:spatial_bias_maps}
\end{figure}
\clearpage
\newpage
\section{Supplemental Information}

\subsection{CMIP Models Used in Figure~\ref{fig:cmip_ensemble}}

The CMIP model simulations used to create Figure~\ref{fig:cmip_ensemble} are listed in Table~\ref{table:cmip_table}.

\begin{table}[t]
\centering
\caption{The following lists the CMIP models and the model variants used to create Figure~\ref{fig:cmip_ensemble}.}
\label{table:cmip_table}
\begin{tabular}{ccccc}
\toprule
\textbf{}           & \textbf{Abrupt4xCO2} & \textbf{piControl} \\
\textbf{}           & \textbf{Variant} & \textbf{Variant} \\ 
\textbf{Model}      & \textbf{and Grid Label} & \textbf{and Grid Label} \\ \midrule
ACCESS-CM2               & r1i1p1f1\_gn                 & r1i1p1f1\_gn               \\
CMCC-CM2-SR5             & r1i1p1f1\_gn                 & r1i1p1f1\_gn               \\
CNRM-ESM2-1              & r1i1p1f2\_gr                 & r1i1p1f2\_gr               \\
CanESM5                  & r1i1p1f1\_gn                 & r1i1p1f1\_gn               \\
IPSL-CM6A-LR             & r1i1p1f1\_gr                 & r1i1p1f1\_gr               \\
MIROC6                   & r1i1p1f1\_gn                 & r1i1p1f1\_gn               \\
MPI-ESM1-2-HR            & r1i1p1f1\_gn                 & r1i1p1f1\_gn               \\
NorESM2-MM               & r1i1p1f1\_gn                 & r1i1p1f1\_gn               \\
\bottomrule
\end{tabular}
\end{table}

\subsection{Hyperparameter Tuning}\label{sec:hyperparameter_tuning}

Our goal in this manuscript is to demonstrate the existence-proof of an ML emulator that can simulate fast responses. In designing the columnwise architecture, we test a number of architectural choices.  We show the results of those experiments here, although we note exhaustive hyperparameter tuning is beyond the scope of the manuscript.  However, we clarify that we do not assert that we are presenting the optimal set of hyperparameters for all possible climate use cases.

We use a training-validation-test set paradigm.  The model was trained on simulation years 12 to 53 (inclusive), the same as \cite{Duncan2024}.  The validation set is used to assess the performance of different architectural hyperparameters.  The validation set uses years 54 to 63 to assess the bias of the ACE in 10-year rollouts.  The years 64 to 65 are also used as a validation set to assess the effect of different model hyperparameters on the $4 \times \text{CO}_2$ hydrological cycle.   The out-of-sample test set is years 66 to 73.  Absolutely no information from these years is used during training or model hyperparameter selection.  One takeaway from this train-validation-test split is that we select model hyperparameters based on a 10-year $1 \times \text{CO}_2$ rollout and a 2-year $4 \times \text{CO}_2$ response.  A second takeaway is that no model training or hyperparameter tuning whatsoever is done with $0.5 \times \text{CO}_2$, $2 \times \text{CO}_2$, or $8 \times \text{CO}_2$, using any data in any of the train, validation, or test sets. The response to these perturbation levels emerges organically with no hyperparameter tuning.

\subsubsection{Adding in Moisture and Dry Air Constraints to the Default ACE}

\subsubsection*{Methods}

The ACE architecture used in \cite{Duncan2024} does not include constraints to conserve moisture and dry air. These constraints were subsequently introduced in ACE2 \citep{Watt-meyer2025}, a newer-generation version of ACE.  These constraints are implemented during training and serve as correction layers on the output of the model.  Because they are applied during training, all of the model weights are updated assuming these constraints exist.  The full ACE constraints are described in detail in \cite{Watt-meyer2025}.  We reproduce them here for convenience.

Dry air is conserved in ACE by setting
\begin{equation}
p_s'(t)
=
p_s(t)
-
\left\langle
p_s^{\mathrm{dry}}(t)
-
p_s^{\mathrm{dry}}(t-1)
\right\rangle
\end{equation}

\noindent where $p_s'(t)$ is the corrected surface pressure tendency, $p_s(t)$ is the surface pressure predicted by the model, and $\langle \ldots \rangle$ denotes the global mean.

The moisture conservation equation is

\begin{eqnarray}~\label{eqn:moisture_conservation}
\frac{\mathrm{TWP}(t+\Delta t) - \mathrm{TWP}(t)}{\Delta t}
& = & E(t) - P(t) + \left.\frac{\partial \mathrm{TWP}}{\partial t}\right|_{\mathrm{adv}} (t)\\
\mathrm{TWP}(t) & = & \frac{1}{g}\int_0^{p_s} q(t,p)\text{d}p \notag \\
\mathrm{LHFLX}(t) &=& E(t)  L_v \notag
\end{eqnarray}

\noindent where TWP is the total water path, $E$ is evaporation, $P$ is precipitation, $t$ is time, the subscript ``adv'' denotes advection within the atmosphere,  $q$ is the specific total water (water vapor + liquid condensate + ice condensate), $g$ is the acceleration of gravity, and $\mathrm{LHFLX}$ is the latent heat flux.   We will treat $\mathrm{LHFLX}$ as interchangeable with $E$ throughout the remainder of this Supplemental Information. By construction, 

\begin{equation}
\left\langle
\left.\frac{\partial \mathrm{TWP}}{\partial t}\right|_{\mathrm{adv}} (t)
\right\rangle = 0
\end{equation}

\noindent Taking the global mean of Equation~\ref{eqn:moisture_conservation} yields

\begin{equation}\label{eqn:global_mean_moisture_conservation}
\left\langle
\frac{\mathrm{TWP}(t+\Delta t) - \mathrm{TWP}(t)}{\Delta t}
\right\rangle
=
\left\langle E(t) - P(t) \right\rangle
\end{equation}

\noindent One way to ensure moisture conservation is to rescale precipitation such that Equation~\ref{eqn:global_mean_moisture_conservation} holds true via

\begin{equation}\label{eqn:moisture_rescale_precip}
P'(t)
=
\frac{P(t)}{\langle P(t) \rangle}
\left\langle
E(t) - \frac{\mathrm{TWP}(t) - \mathrm{TWP}(t-1)}{\Delta t}
\right\rangle
\end{equation}

\noindent where $P'(t)$ is the corrected precipitation rate at time $t$ and $P(t)$ is the precipitation prior to this correction.

\newpage
Alternately, another way to ensure moisture conservation is to rescale evaporation such that Equation~\ref{eqn:global_mean_moisture_conservation} holds true.

\begin{equation}\label{eqn:moisture_rescale_evap}
E'(t)
=
\frac{E(t)}{\langle E(t) \rangle}
\left\langle
P(t) + \frac{\mathrm{TWP}(t) - \mathrm{TWP}(t-1)}{\Delta t}
\right\rangle
\end{equation}

\noindent where $E'(t)$ is the corrected evaporation rate at time $t$ and $E(t)$ is the evaporation prior to this correction.

\begin{figure}
    \centering
    \includegraphics[width=\linewidth]{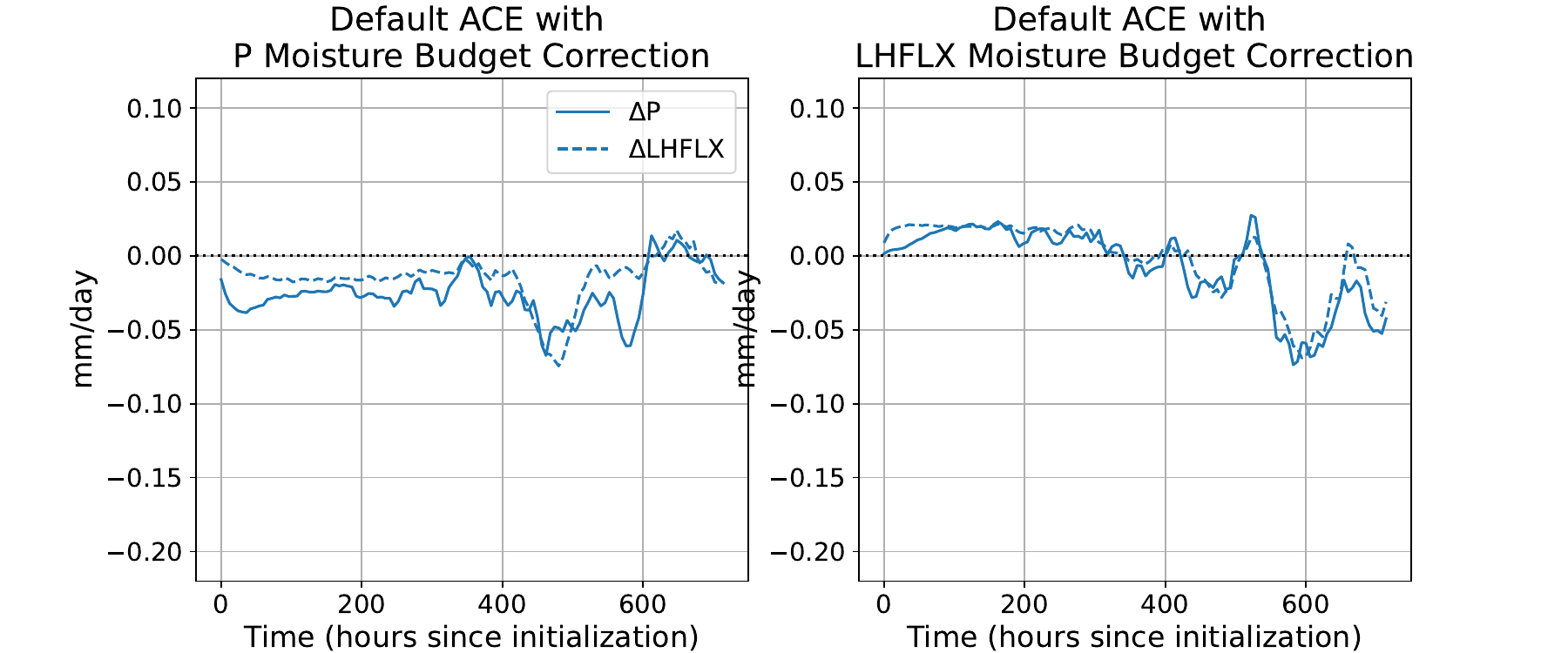}
    \caption{\textbf{Fast Responses in the Default ACE Architecture with Moisture Constraints}.  The response of precipitation (``P'') and latent heat flux(``LHFLX'') to $4\times$\cotwo\ is shown in the default ACE architecture with moisture and dry air conservation \cite{Watt-meyer2025} added during training.  Left (right) panels: Moisture conservation imposed by Equations~\ref{eqn:moisture_rescale_precip} (\ref{eqn:moisture_rescale_evap}). The response is calculated only using the two years in the validation set (simulation years 0064-0065).}
    \label{fig:moisture_constraint_hyperparameter}
\end{figure}

We first test the effect of applying these moisture conservation methods to the default ACE architecture (without the columnwise diagnostic neural networks).  
We train a new ACE checkpoint with the default architecture with moisture constraints added in.  For this architecture, Figure~\ref{fig:moisture_constraint_hyperparameter} shows that the magnitude of the global mean response to 4$\times$\cotwo\ is incorrect relative to the response obtained in the benchmark calculations performed with the EAMv2 (Figure~\ref{fig:global_mean_response}a). In addition, the evaporation moisture constraint yields a response with the wrong sign. Therefore, we find that these constraints alone are not sufficient to obtain the correct fast response.  Even though moisture constraints alone are not enough to correctly simulate fast responses, we still add them into the columnwise ACE presented in the main text of this paper, as it is intrinsically advantageous to have a model with built-in moisture conservation.

\subsubsection{Effect of Joint Training}

\begin{figure}
    \centering
    \includegraphics[width=\linewidth]{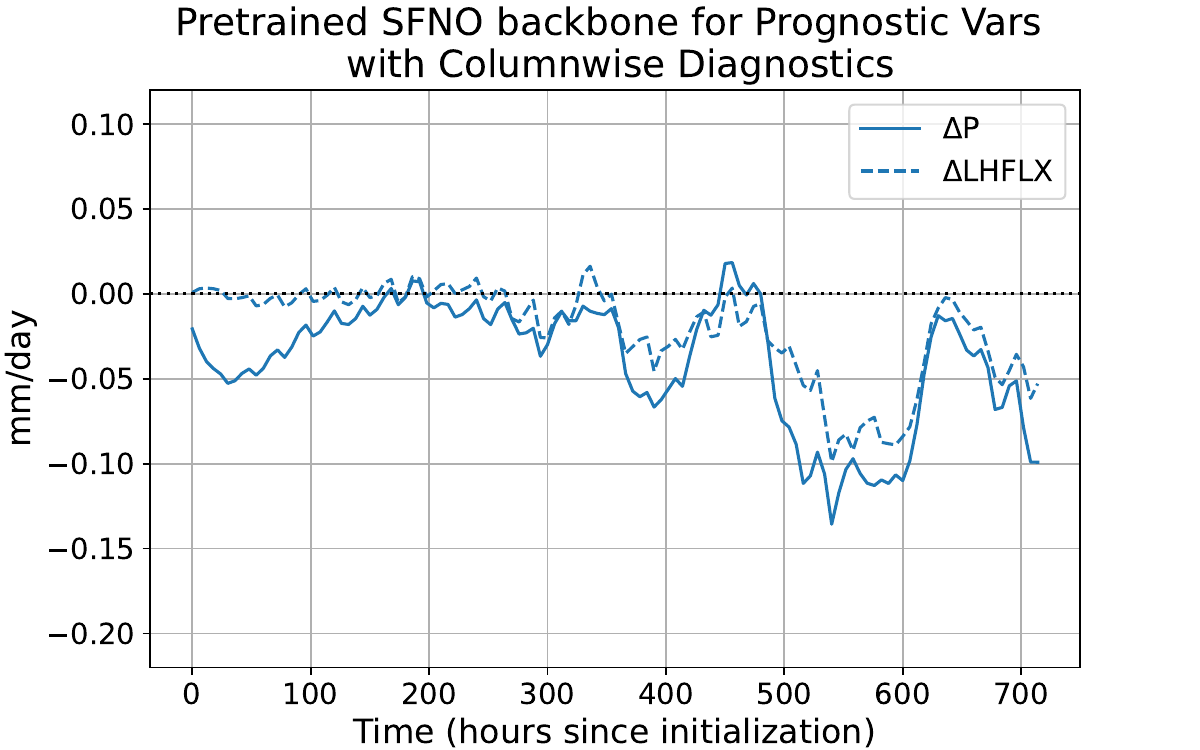}
    \caption{\textbf{Fast Responses in the Columnwise ACE Architecture with SFNO and the Column Diagnostic Network Trained Separately}.  The response of precipitation (``P'') and latent heat flux (``LHFLX'') to $4 \times\text{CO}_2$ is shown in a columnwise ACE architecture with a modified training setup.  First, SFNO is trained for the prognostic variables.  Then, the weights of SFNO are frozen and the column diagnostic network is trained. The fast response is calculated only using the two years in the validation set (simulation years 0064-0065).}
    \label{fig:pretrained_SFNO}
\end{figure}

In the default ACE, SFNO predicts both the prognostic and diagnostic variables.  In the columnwise ACE architecture, SFNO predicts the prognostic variables, and a separate columnwise neural network predicts the diagnostic variables. For the chosen network presented in the main text of this paper, we train the SFNO and the columnwise neural network jointly from scratch.  In this section, we explore the effect of training in multiple stages.  First, we train SFNO on prognostic variables: this is the network that performs the timestepping.  Then, we freeze the SFNO weights and train the diagnostic models to predict additional variables at a given time.  This workflow has been proposed in previous literature \cite{Pathak2022, Mitra2023}, and it has been discussed as a way to add new variables to a network without retraining a new prognostic timestepping backbone from scratch. In Figure~\ref{fig:pretrained_SFNO}, we show the fast response of a model trained with this multistage approach.  The magnitude of the response is still too small compared to the expected response of EAMv2 in the same years on the validation set (Figure~\ref{fig:global_mean_response}a).

\subsubsection{Spherical Convolutions vs. Planar Convolutions}

\begin{figure}
    \centering
    \includegraphics[width=0.5\textheight]{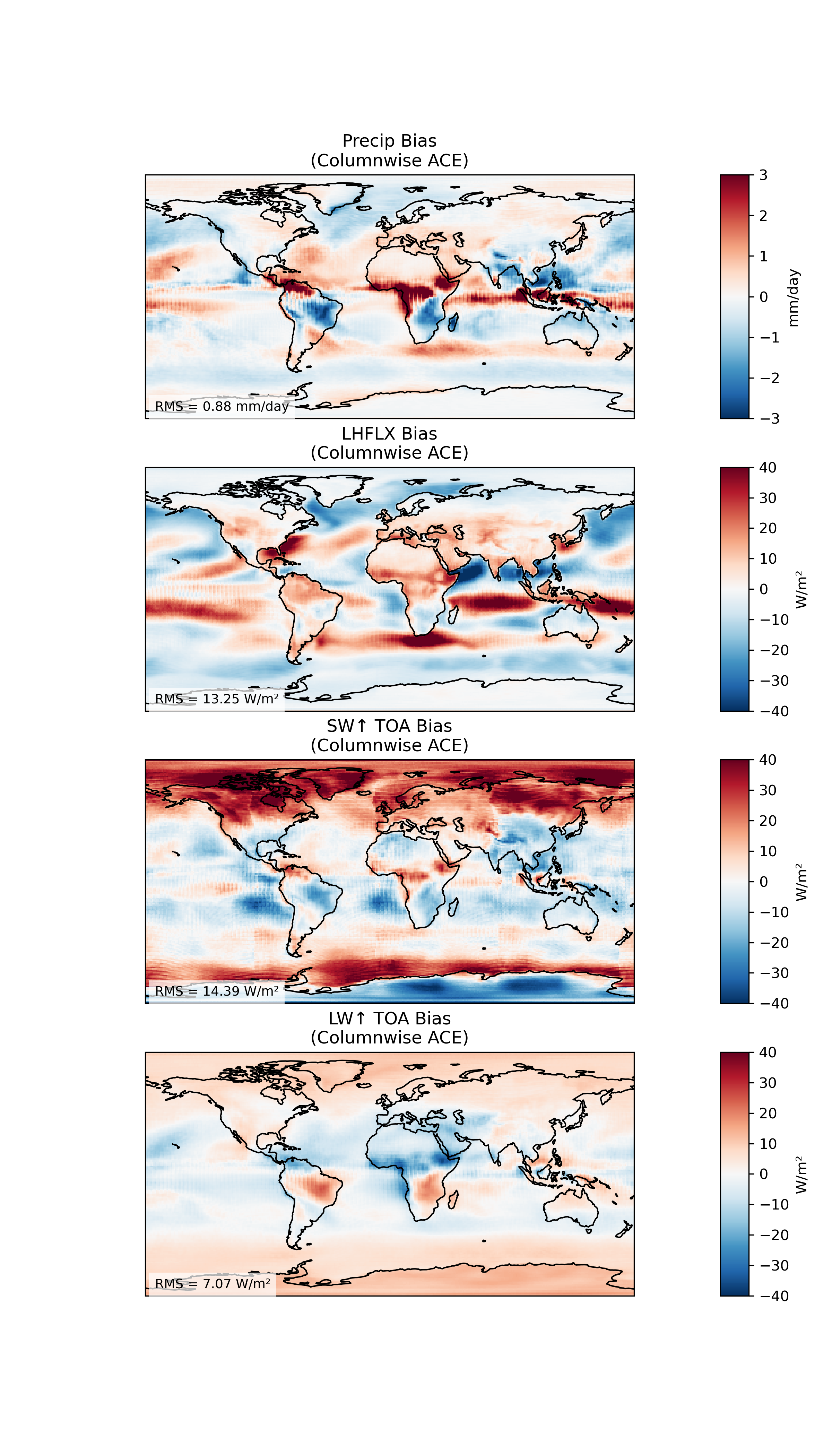}
    \caption{\textbf{10-year Bias using Planar Convolutions for Columnwise Diagnostics.} Same as Figure~\ref{fig:spatial_bias_maps} but for a model that uses planar $3 \times 3$ diagnostics.}
    \label{fig:planar_bias}
\end{figure}

Spherical convolutions preserve spherical geometry when calculating the $3\times3$ convolution to account for neighboring columns.  On the other hand, standard planar convolutions do not preserve this geometry.  The columnwise ACE with planar convolutions has significantly more bias on 10-year rollouts (Figure~\ref{fig:planar_bias}).  This model architecture thus fails criteria \#3 in Section~4.4 on requirements for a successful emulator of fast responses.

\subsubsection{Column Timestepping}

\begin{figure}
    \centering
    \includegraphics[width=\linewidth]{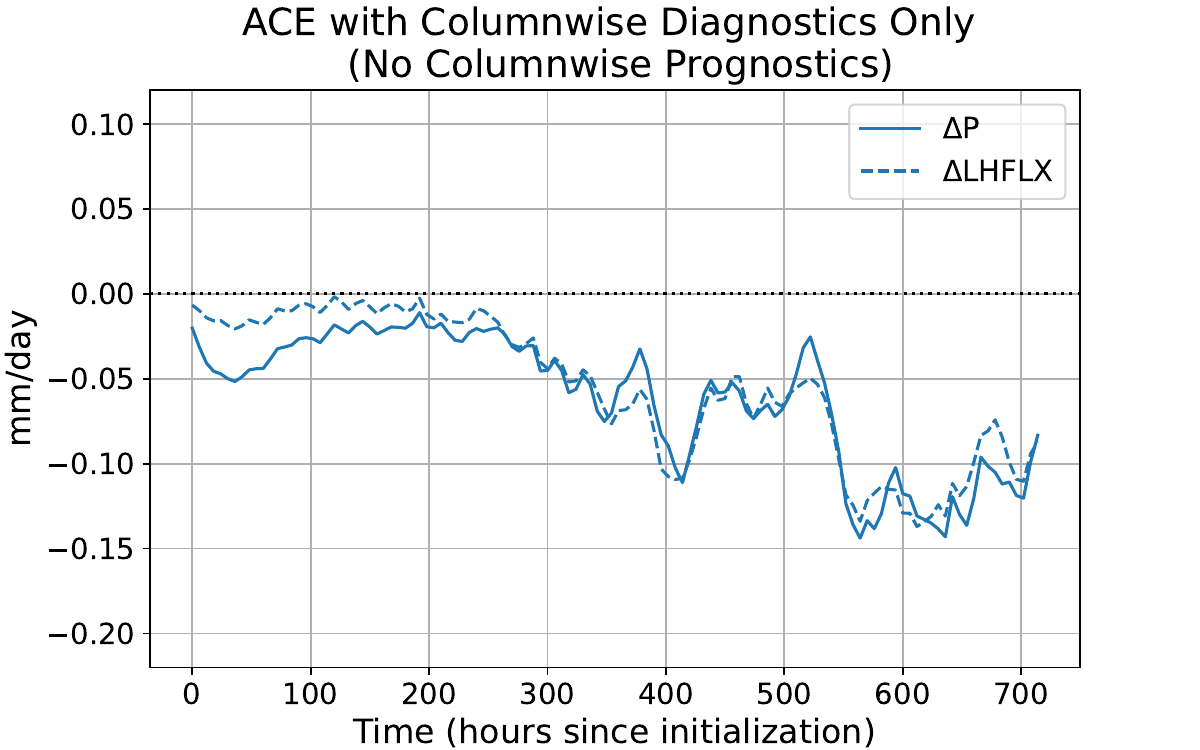}
    \caption{\textbf{Fast Responses in a Columnwise ACE Architecture with Column Diagnostics but No Column Prognostics}.  The response of precipitation (``P'') and latent heat flux (``LHFLX'') to $4 \times\text{CO}_2$ is shown in a columnwise ACE architecture with a column diagnostic network but no column prognostic network.  See Figure~\ref{fig:architecture_diagram} for a description of these two networks. The fast response is calculated only using the two years in the validation set (simulation years 0064-0065).}
    \label{fig:nocolumntimestep_performance}
\end{figure}

The columnwise timestepping is described in Methods Section 4.6.2 and Figure~\ref{fig:architecture_diagram}.  With columnwise  timestepping, the prognostic variables are calculated as the sum of a term from SFNO and a term from a column-local neural network.  Without columnwise timestepping, the response to the $4 \times \text{CO}_2$ perturbation degrades.  Figure~\ref{fig:nocolumntimestep_performance} shows that the precipitation and latent heat flux response of the architecture is muted.  The expected response nears 0.11 mm/day after about 300 days since initialization (Figure~\ref{fig:global_mean_response}), but this architecture's response is only about 0.02 mm/day on this time scale.

\subsubsection{Position Embedding in Diagnostic Model and Prognostic Model}

The default SFNO structure has three major parts: the encoder, the Fourier neural operator blocks, and the decoder.  The encoder maps the input atmospheric state into a latent space, with a specified ``embedding dimension," which determines the size of this latent space.   The Fourier neural operator blocks apply learned spectral filters.  Finally, the decoder maps the input from the latent space back to the physical atmospheric state.  

Latitude and longitude are not included directly as input variables.  However, the SFNO can learn location-specific information through its ``position embedding." The position embedding is a learnable parameter that is added to each grid cell in the latent space of SFNO.  This position embedding is entirely learned during training, with no pre-specified values or inputs.  We find empirically that position embeddings in ACE help with stability in long rollouts.

\begin{figure}
    \centering
    \includegraphics[width=\linewidth]{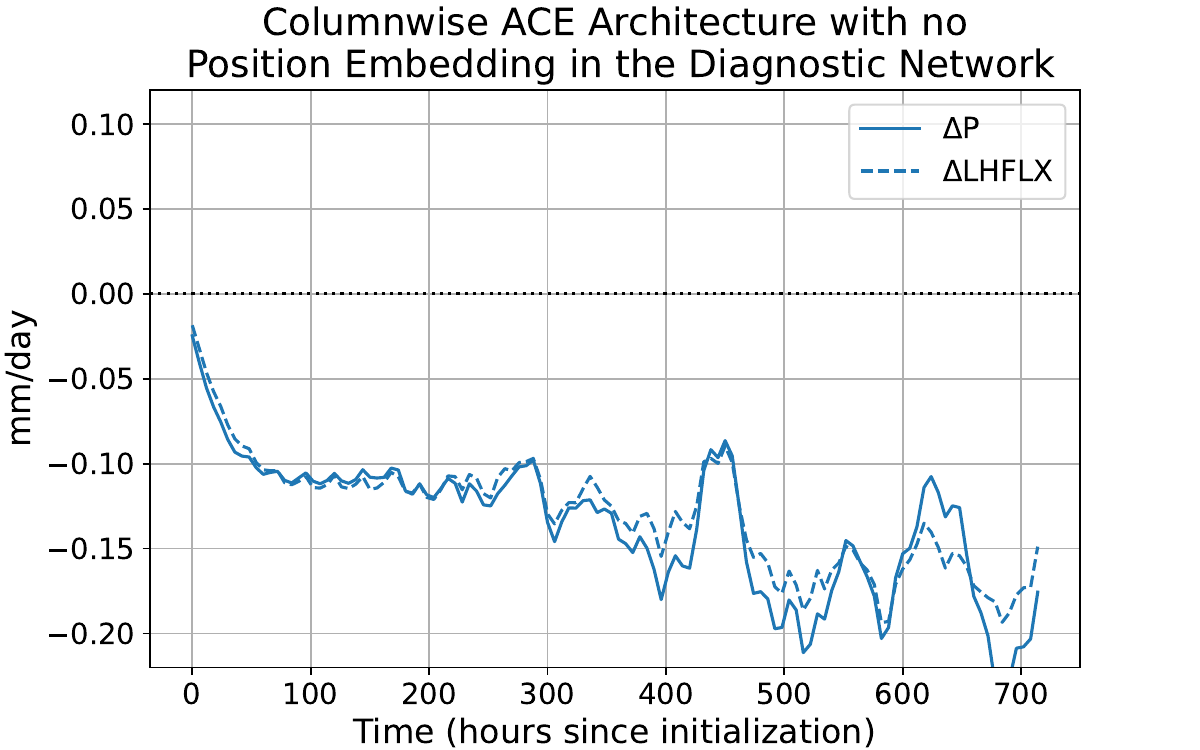}
    \caption{\textbf{Fast Responses in a Columnwise ACE Architecture with No Position Embedding for Diagnostics}.  The response of precipitation (``P'') and latent heat flux (``LHFLX'') to $4 \times\text{CO}_2$ is shown in a columnwise ACE architecture with no position embedding used in the diagnostic network.  The position embedding is a learnable vector added to each grid cell in the latent space of ACE.  The fast response is calculated only using the two years in the validation set (simulation years 0064-0065).}
    \label{fig:coldiagnostics}
\end{figure}

In the columnwise ACE, a separate neural network is used to predict the diagnostic variables.  To be as consistent with the default ACE architecture as possible, we add a position embedding to the column diagnostic network used in the main text of this paper.  (It uses its own position embedding and does not reuse the one from the SFNO.)  However, we also train a columnwise ACE with no position embedding in the column diagnostic network.  We find that this model also exhibits promising performance on fast responses (Figure~\ref{fig:coldiagnostics}).  We suggest that the use of position embeddings for prognostic variables and diagnostic variables should be an area of future research and ML model development.

\subsubsection{Results From Purely Local Column Training}

\begin{figure}
    \centering
    \includegraphics[width=0.5\textheight]{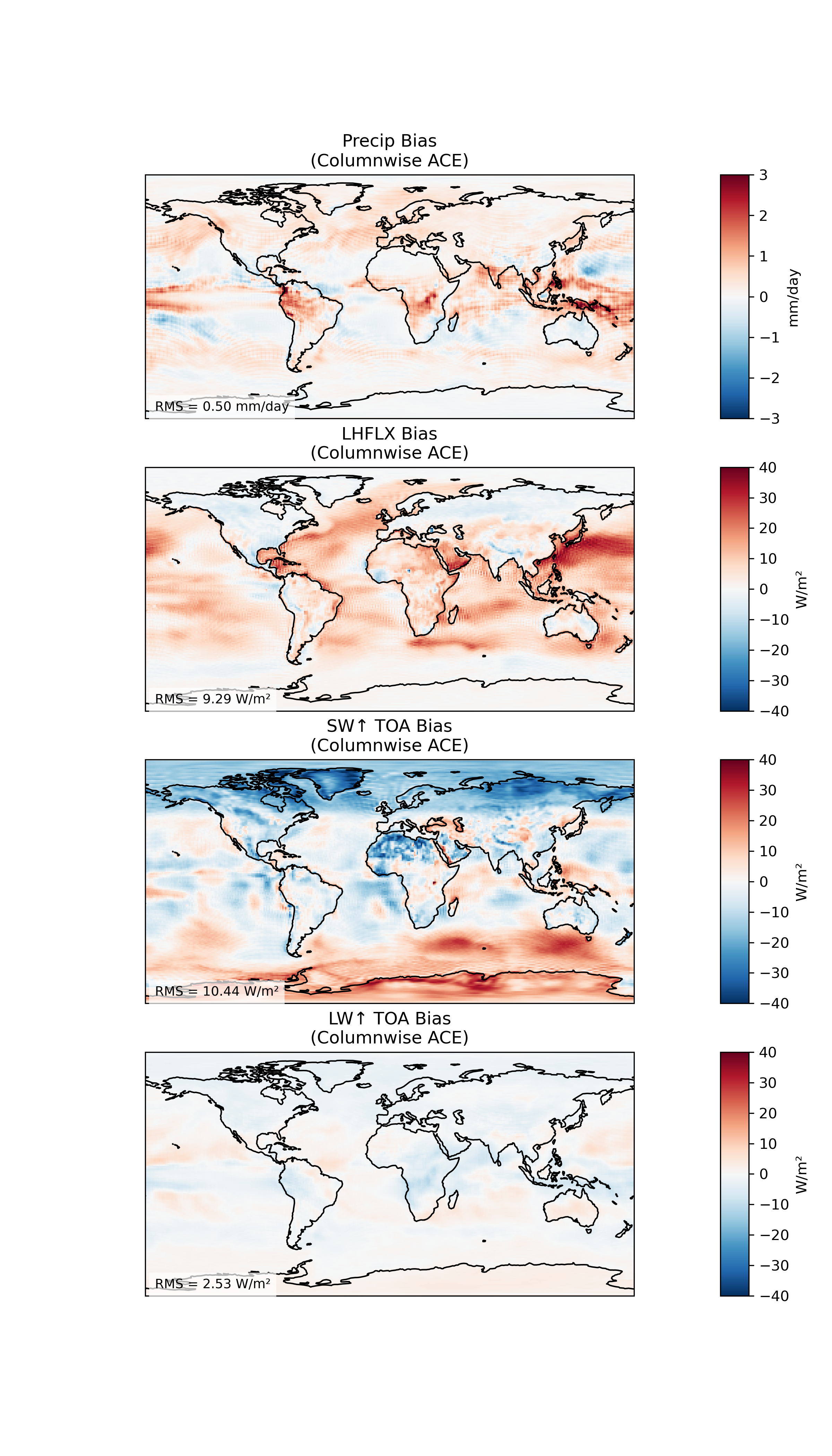}
    \caption{\textbf{10-year Bias using Purely Local Column Diagnostics.} The diagnostic architecture presented in Figure~\ref{fig:spatial_bias_maps} uses a $3 \times 3$ convolution kernel in its first layer, which implies that the diagnostic variables are predicted using a $3 \times 3$ window.  This figure shows the bias maps of a 10-year rollout on the validation set, with no nonlocal window. }
    \label{fig:spatial_bias_maps_purely_local}
\end{figure}

In our columnwise ACE, the column diagnostic network predicts the diagnostics using a $3 \times 3$ convolution in its first layer (the subsequent layers do not use any local convolutions).  For the first layer, we use $3 \times 3$ spherical convolutions to allow the model to access the vertical profiles from nearby grid cells when diagnosing precipitation.  Since ACE has a long timestep (6 hours), access to the nearby grid cells allows the model to maintain low bias (Figure~\ref{fig:spatial_bias_maps}).  If the model is purely local, then its bias on 10-year rollouts is higher than the bias of the default ACE (Figure~\ref{fig:spatial_bias_maps_purely_local}.)

\subsubsection{Motivation for Columnwise Architecture from NeuralGCM}

\begin{figure}
    \centering
    \includegraphics[width=\linewidth]{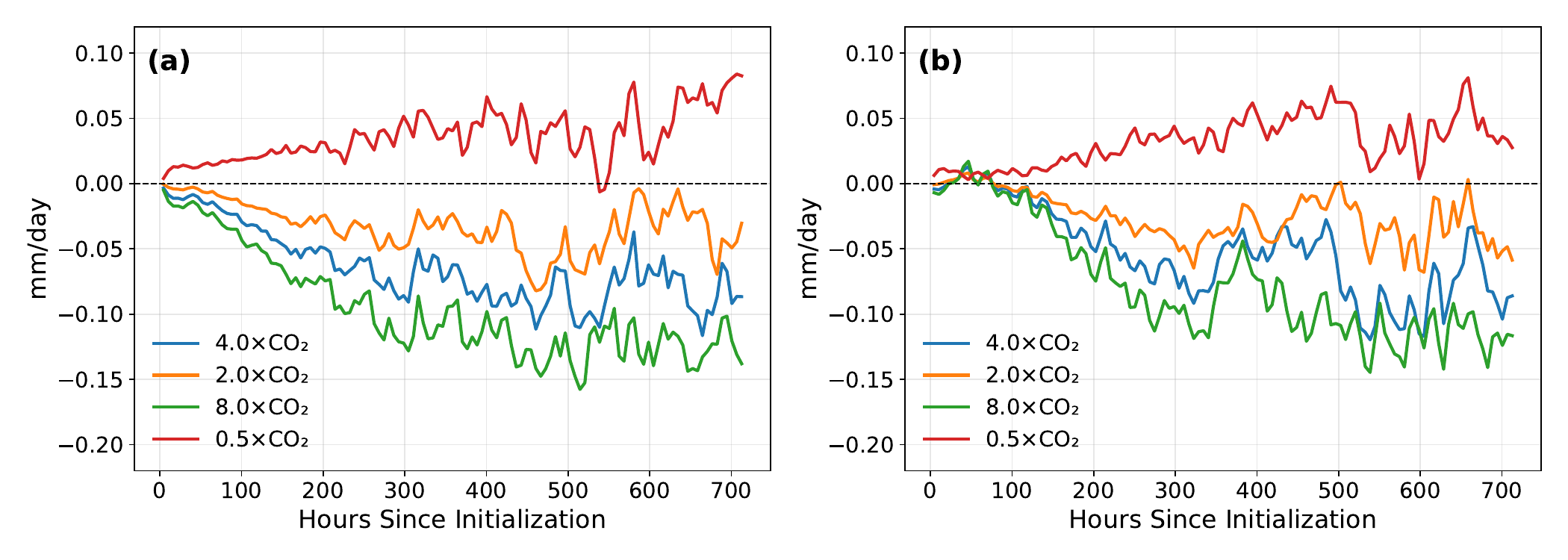}
    \caption{\textbf{Precipitation Response to Increased \cotwo\ in NeuralGCM} Panel (a) shows the response from a version of NeuralGCM that predicts precipitation and diagnoses evaporation as a residual.  Panel (b) shows the response from the other version of NeuralGCM, which predicts evaporation and diagnoses precipitation as a residual. }
    \label{fig:neuralgcm_results}
\end{figure}

Our motivation for a columnwise ACE architecture is based on full-physics GCMs \cite{Balaji2022} and NeuralGCM \cite{Kochkov2024, Yuval2026}, both of which use a column abstraction.  NeuralGCM is a hybrid model: it combines a physics-based dynamical core with a column-local ML model for parameterized processes.  NeuralGCM also predicts precipitation using a column-local network \cite{Yuval2026}.  We find that NeuralGCM has the correct sign of the hydrological cycle's response to \cotwo, unlike the default ACE (Figure~\ref{fig:neuralgcm_results}).  However, we are unable to perform additional, in-depth validation of NeuralGCM for this task because it is trained on ERA5.  As ERA5 is a reanalysis product, there is no way to quadruple \cotwo\ and thus compare NeuralGCM to an estimate of the ``true" response, as we do with ACE and EAM in Figures~\ref{fig:global_mean_response} and \ref{fig:spatial_patterns}.  We can only verify the sign and approximate logarithmic relation between \cotwo\ levels.  Even full-physics GCMs disagree on the exact magnitude of the hydrological response \cite{Stjern2023}. We open-source the code to couple RRTMG to NeuralGCM to catalyze future development in this area and to facilitate tests with future versions of NeuralGCM.

A caveat to the NeuralGCM test is that NeuralGCM does not output surface temperature over land, which is a necessary input to RRTMG.  Therefore, we use climatological land surface temperatures as input to RRTMG to calculate the $N \times \text{CO}_2 - 1 \times \text{CO}_2$ perturbations for this proof-of-principle test.

\end{appendices}


\clearpage
\bibliography{sn-bibliography}

\end{document}